\documentclass{article}
\usepackage{graphicx}
\usepackage{amsmath}
\usepackage{amsfonts}
\usepackage{amssymb}

\begin{document}

\title{Density functionals derived from Feynman diagrams. }
\author{Georges Ripka
\and ECT*, Villa Tambosi, I-38050 Villazano (Trento), ITALY
\and Institut de Physique Theorique,
\and Centre d'Etudes de Saclay, F-91191
\and Gif-sur-Yvette Cedex, FRANCE
\and ripka@cea.fr}
\maketitle
\begin{abstract}
We construct a stationary density functional for the partition function from a
chosen set of one (boson) line irreducible Feynman diagrams. The construction
does not proceed by the inversion of a Legendre transform. It is formulated
for fermions with Coulomb interactions, in which case a stationary 
functional of the particle density is contructed, as well as for nucleons 
interacting with mesons, which involves a stationary functional of 
several densities. The use of Kohn-Sham orbits is shown to be an 
unnecessary complication.
\end{abstract}

\section{Introduction.}
\setcounter{equation}{0}\renewcommand{\theequation}{\arabic{section}%
.\arabic{equation}}

The advantage of constructing density functionals of finite systems, such as
nuclei, atoms and crystals, resides in the fact that they give rise to single
particle states which can be calculated with a static (energy independent)
potential.\ This makes computations considerably easier if one needs to go
beyond the mean field approximation. For systems which are composed, for
example, of electrons with a Coulomb interaction, it is natural to construct a
functional of the particle density $n\left(  \vec{r}\right)  =\left\langle
\psi^{\dagger}\left(  \vec{r}\right)  \psi\left(  \vec{r}\right)
\right\rangle $ because this is the density to which the Coulomb field
couples.\ It is for such systems that density functionals were first
constructed\cite{KohnHohenberg1964,KohnSham1965}. In nuclei however, there is
no reason to single out the particle density $n\left(  \vec{r}\right)  $.\ For
example, pions couple to the density $\left\langle \psi^{\dagger}\left(
\vec{r}\right)  \tau_{a}\gamma_{5}\vec{\alpha}\psi\left(  \vec{r}\right)
\right\rangle $, scalar and vector mesons couple to densities $\left\langle
\psi^{\dagger}\left(  \vec{r}\right)  \gamma_{0}\psi\left(  \vec{r}\right)
\right\rangle $ and $\left\langle \psi^{\dagger}\left(  \vec{r}\right)
\psi\left(  \vec{r}\right)  \right\rangle $ respectively.\ In this work, we
show how to express the partition function of the finite system in terms of a stationary
functional of densities which are determined by the nature of the interactions.

The construction of stationary functionals of the density matrix, appropriate
to systems with two-body interactions, was developed in Ref.\cite{Ripka86},
based on an earlier work \cite{Ripka70} which described the theory of a static
potential in the presence of two-particle two-hole excitations. 
In these works, the density
functionals are derived from a chosen set of one (interaction) line
irreducible Feynman or Goldstone diagrams. More recently, considerable work
has been devoted to the construction of stationary density functionals
obtained from a Legendre transform (see \cite{Fukuda1994}%
,\cite{Fukuda1995}, \cite{Okumura1996},\cite{ValievFernando1998},
\cite{Polonyi2004},\cite{Furnstahl2009-1} and references therein).

In this work, we construct a density functional from a 
chosen subset of one line
irreducible Feynman diagrams which are assumed to be dominant. We consider
both fermions with Coulomb interactions and nucleons interacting with scalar,
vector and $\pi$ mesons. The formalism is simpler in the former case because
it involves only one single-particle density, 
whereas several densities are involved in
the latter. The partition functions of the two systems are defined in Sections
\ref{sec:fermsclb} and \ref{sec:nucmesons}. The Feynman diagram rules for the
two systems are summarized in Appendices \ref{ap:diagclb} and
\ref{ap:nucmesdiag}. Section \ref{sec:irrdiag} describes the properties of
one-line irreducible diagrams. In section \ref{sec:selfconsist} the single
particle densities are expressed in terms of one-line irreducible diagrams
which are calculated in terms of self-consistent potentials. The potentials
and single particle densities are related by a simple analytic expression.
Section \ref{sec:iterate} describes iteration procedures to calculate the
self-consistent potentials and the single particle densities. In Section
\ref{sec:partirr} the partition function is expressed in terms of one-line
irreducible diagrams and in Section \ref{sec:stat} it is shown to be a
stationary functional of either the potentials or of the single particle
densities. In Section \ref{sec:kohnsham} the theory is expressed in terms of
so-called Kohn-Sham orbits the use of which is shown to be an unnecessary complication.

\section{Non-relativistic fermions with Coulomb interactions.}
\label{sec:fermsclb}
\setcounter{equation}{0}\renewcommand{\theequation}{\arabic{section}%
.\arabic{equation}}

Non-relativistic fermions, such as electrons with Coulomb interactions, can be
described by a hamiltonian of the form:%
\[
H-\mu N=\int d^{3}r\;\psi^{\dagger}\left(  \vec{r}\right)  \left(
-\frac{\nabla^{2}}{2M}+U_{ext}\left(  \vec{r}\right)  -\mu\right)  \psi\left(
\vec{r}\right)
\]%
\begin{equation}
+\frac{1}{2}\int d^{3}r_{1}d^{3}r_{2}\;\psi^{\dagger}\left(  \vec{r}%
_{1}\right)  \psi^{\dagger}\left(  \vec{r}_{2}\right)  \frac{e^{2}}%
{4\pi\left|  \vec{r}_{1}-\vec{r}_{2}\right|  }\psi\left(  \vec{r}_{2}\right)
\psi\left(  \vec{r}_{1}\right)  \label{hann}%
\end{equation}
where $\psi$ is the fermion field, $M$ the fermion mass, $e$ its electric
charge and $U_{ext}\left(  \vec{r}\right)  $ an external potential which could
be due, for example, to an applied external field, to a nucleus or to the
crystal nuclei. We define:%
\begin{equation}
h_{0}=-\frac{\nabla^{2}}{2M}+U_{ext}\left(  \vec{r}\right)  \label{hnonrel}%
\end{equation}
The partition function is given by:%
\begin{equation}
Z=Tr\;e^{-\beta\left(  H-\mu N\right)  }\equiv e^{W}=\int D\left(
\psi^{\dagger},\psi\right)  e^{-I\left(  \psi^{\dagger},\psi\right)
}\label{partclbpsi}%
\end{equation}
where $\psi^{\dagger}$ and $\psi$ are independent Grassmann variables and
where $I\left(  \psi^{\dagger},\psi\right)  $ is the \emph{euclidean} action:%
\[
I\left(  \psi^{\dagger},\psi\right)  =
\]%
\begin{equation}
\int d^{4}x\left[  \psi^{\dagger}\left(  x\right)  \left(  \partial_{\tau
}+h_{0}-\mu\right)  \psi\left(  x\right)  -\frac{1}{2}\int d^{4}x_{1}%
d^{4}x_{2}\;\psi^{\dagger}\left(  x_{1}\right)  \psi\left(  x_{1}\right)
\left\langle x_{1}\left|  K\right|  x_{2}\right\rangle \psi^{\dagger}\left(
x_{2}\right)  \psi\left(  x_{2}\right)  \right]  \label{actnn}%
\end{equation}
with $x_{\mu}=x^{\mu}=\left(  \tau,\vec{r}\right)  $ and:%
\begin{equation}
-K=\frac{e^{2}}{-\nabla^{2}}\;\;\;\;\;\;\;-\left\langle x_{1}\left|  K\right|
x_{2}\right\rangle =\delta\left(  \tau_{1}-\tau_{2}\right)  \frac{e^{2}}%
{4\pi\left|  \vec{r}_{1}-\vec{r}_{2}\right|  }\label{coulpot}%
\end{equation}
Note that we defined the Coulomb potential to be $-K$ (and not $K$) so that
$K$ is a negative definite operator. The reason is to use a common notation
for fermions with Coulomb interactions and nucleons interacting with mesons,
described in Section \ref{sec:nucmesons}. We can also express the partition
function in terms of an integral over an additional boson field $\omega_{0}$:%
\[
Tr\;e^{-\beta\left(  H-\mu N\right)  }=e^{W}=\int D\left(  \psi^{\dagger}%
,\psi\right)  D\left(  \omega_{0}\right)  e^{-I\left(  \psi^{\dagger}%
,\psi,\omega_{0}\right)  }%
\]%
\begin{equation}
I\left(  \psi^{\dagger},\psi,\omega_{0}\right)  =\int d^{4}x\;\psi^{\dagger
}\left(  x\right)  \left(  h_{0}-\mu+i\omega_{0}\right)  \psi\left(  x\right)
-\frac{1}{2}\int d^{4}x_{1}d^{4}x_{2}\;\omega_{0}\left(  x_{1}\right)
\left\langle x_{1}\left|  K^{-1}\right|  x_{2}\right\rangle \omega_{0}\left(
x_{2}\right)  \label{partclb}%
\end{equation}
By integrating over the field $\omega_{0}$ we recover the action
(\ref{actnn}).\ By integrating over the nucleon fields $\psi^{\dagger}$ and
$\psi$ we obtain the partition function in the form:%
\[
Z=Tre^{-\beta\left(  H-\mu N\right)  }\equiv e^{W}=\int D\left(  \omega
_{0}\right)  e^{-I\left(  \omega_{0}\right)  }%
\]%
\begin{equation}
I\left(  \omega_{0}\right)  =-Tr\ln\left(  \partial_{\tau}+h_{0}-\mu
+i\omega_{0}\right)  -\frac{1}{2}\int d^{4}x_{1}d^{4}x_{2}\;\omega_{0}\left(
x_{1}\right)  \left\langle x_{1}\left|  K^{-1}\right|  x_{2}\right\rangle
\omega_{0}\left(  x_{2}\right)  \label{actclb}%
\end{equation}
In (\ref{partclb}) the trace is taken over the Hilbert space composed of
$0,1,2,...$ fermions.\ In equation (\ref{actclb}) and in the following, $Tr$
is a trace over the euclidean space-time $\left(  \tau,\vec{r}\right)  $ and
over the spins of the fermions, in other words, over the arguments of the
fermion field $\psi$:%
\begin{equation}
Tr\;A=tr\int d^{4}x\;\left\langle x\left|  A\right|  x\right\rangle =\int
_{0}^{\beta}d\tau\int d^{3}r\sum_{\sigma}\left\langle \tau\vec{r}\sigma\left|
A\right|  \tau\vec{r}\sigma\right\rangle  \label{trace1}
\end{equation}
and $tr$ denotes a trace over all the discrete indices other than $\left(
\tau,\vec{r}\right)  $:%
\begin{equation}
tr\;A=\sum_{\sigma}\left\langle \sigma\left|  A\right|  \sigma\right\rangle  \label{trace2}
\end{equation}
The chemical potential of the fermions is $\mu$. In the zero temperature limit
$\beta\rightarrow\infty$, the energy $E$ of the system composed of $N$
fermions is given by:%
\begin{equation}
E-\mu N=-\frac{1}{\beta}W
\end{equation}

The reader who is not interested in nucleons interacting with mesons can
proceed to Section \ref{sec:irrdiag}.

\section{Nucleons interacting with mesons.}
\label{sec:nucmesons}
\setcounter{equation}{0}\renewcommand{\theequation}{\arabic{section}%
.\arabic{equation}}

We consider a system of nucleons of mass $M$ interacting with a scalar meson
$\sigma$ of mass $m_{\sigma}$, an isoscalar vector meson $\omega_{\mu}$ of
mass $m_{\omega}$ and pions $\pi_{a}$ of mass $m_{\pi}$. The partition
function can be expressed as the following path integral over the nucleon and
meson fields:
\begin{equation}
Tre^{-\beta\left(  H-\mu N\right)  }\equiv e^{W}=\int D\left(  N^{\dagger
},N\right)  D\left(  \sigma\right)  D\left(  \omega\right)  D\left(
\pi\right)  e^{-I\left(  N^{\dagger},N,\sigma,\omega,\pi\right)
}\label{partmes1}%
\end{equation}
where $I\left(  N^{\dagger},N,\sigma,\pi,\omega\right)  $ is the
\emph{euclidean} action:%
\[
I_{j}\left(  N^{\dagger},N,\sigma,\pi,\omega\right)
\]%
\[
=N^{\dagger}\left(  \partial_{\tau}+\frac{\vec{\alpha}\cdot\vec{\nabla}}%
{i}+\beta M-\mu+g\beta\sigma+\frac{g_{A}}{2f_{\pi}}\beta\tau_{a}\gamma
_{5}\gamma_{\mu}\left(  \partial_{\mu}\pi_{a}\right)  +\frac{1}{4f_{\pi}^{2}%
}\beta\varepsilon_{abc}\tau_{c}\pi_{a}\gamma_{\mu}\left(  \partial_{\mu}%
\pi_{b}\right)  +g_{\omega}\beta\gamma_{\mu}\omega_{\mu}\right)  N\;
\]%
\begin{equation}
+\int d^{4}x\left[  +\frac{1}{2}\sigma\left(  -\partial^{2}+m_{\sigma}%
^{2}\right)  \sigma+\frac{1}{2}\pi_{a}\left(  -\partial^{2}+m_{\pi}%
^{2}\right)  \pi_{a}+\frac{1}{2}\omega_{\mu}\left(  \left(  -\partial
^{2}+m_{\omega}^{2}\right)  \delta_{\mu\nu}+\partial_{\mu}\partial_{\nu
}\right)  \omega_{\nu}\right]  \label{act}%
\end{equation}
The euclidean action (\ref{act}) is expressed in terms of the following
euclidean $\left(  g_{\mu\nu}=\delta_{\mu\nu}\right)  $ 4-vectors:%
\[
x_{\mu}=x^{\mu}=\left(  \tau,\vec{r}\right)  \;\;\;\;\;\;\gamma_{\mu}%
=\gamma^{\mu}=\left(  i\beta,\vec{\gamma}\right)  \;\;\;\;\;\;\omega_{\mu
}=\omega^{\mu}=\left(  \omega_{0},\vec{\omega}\right)
\]%
\begin{equation}
\partial^{2}=\partial_{\tau}^{2}+\nabla^{2}\;\;\;\;\;\;\;\int d^{4}x=\int
_{0}^{\beta}d\tau\int d^{3}r\label{omegamu}%
\end{equation}
We shall work with the euclidean action obtained by integrating out the
nucleon fields, in which case the partition function becomes:%
\begin{equation}
Tre^{-\beta\left(  H-\mu N\right)  }\equiv e^{W}=\int D\left(  \sigma\right)
D\left(  \omega\right)  D\left(  \pi\right)  e^{-I\left(  \sigma,\omega
,\pi\right)  }\label{partmesnuc}%
\end{equation}
with:%
\[
I\left(  \sigma,\pi,\omega\right)
\]%
\[
=-Tr\ln\left(  \partial_{\tau}+\frac{\vec{\alpha}\cdot\vec{\nabla}}{i}+\beta
M-\mu+g\beta\sigma+\frac{g_{A}}{2f_{\pi}}\beta\tau_{a}\gamma_{5}\gamma_{\mu
}\left(  \partial_{\mu}\pi_{a}\right)  +\frac{1}{4f_{\pi}^{2}}\beta
\varepsilon_{abc}\tau_{c}\pi_{a}\gamma_{\mu}\left(  \partial_{\mu}\pi
_{b}\right)  +g_{\omega}\beta\gamma_{\mu}\omega_{\mu}\right)  \;
\]%
\begin{equation}
+\int d^{4}x\left[  \frac{1}{2}\sigma\left(  -\partial^{2}+m_{\sigma}%
^{2}\right)  \sigma+\frac{1}{2}\pi_{a}\left(  -\partial^{2}+m_{\pi}%
^{2}\right)  \pi_{a}+\frac{1}{2}\omega_{\mu}\left(  \left(  -\partial
^{2}+m_{\omega}^{2}\right)  \delta_{\mu\nu}+\partial_{\mu}\partial_{\nu
}\right)  \omega_{\nu}\right]  \label{act1}%
\end{equation}
The trace $Tr$ is taken over the quantum numbers which define the fermion
field $N\left(  x\right)  $. For example:
\begin{equation}
Tr\left(  \partial_{\tau}+\frac{\vec{\alpha}\cdot\vec{\nabla}}{i}+\beta
M-\mu+g\beta\sigma\right)  \equiv\int d^{4}x\;tr\left\langle x\left|
\partial_{\tau}+\frac{\vec{\alpha}\cdot\vec{\nabla}}{i}+\beta M-\mu
+g\beta\sigma\right|  x\right\rangle \label{trace}%
\end{equation}
and $tr$ is a trace over the discrete (Dirac,flavor) quantum numbers other
than $x=\left(  \tau,\vec{r}\right)  $.

We stress from the outset that we are \emph{not} suggesting that (\ref{act1})
is the action which should be used to calculate finite nuclei. We are simply
introducing the $\sigma,\omega$ and $\pi$ mesons to show how to construct a
density functional with different types of meson-nucleon interactions.

In order to express the partition function in terms of Feynman diagrams, we
adopt a condensed notation. We define the unperturbed fermion hamiltonian:
\begin{equation}
h_{0}=\frac{\vec{\alpha}\cdot\vec{\nabla}}{i}+\beta M \label{hzero}%
\end{equation}
The following is not restricted to a Dirac unperturbed hamiltonian.\ It could,
for example, be replaced by its non-relativistic reduction $h_{0}%
=-\frac{\nabla^{2}}{2M}$ , in which case the couplings would have to be
modified accordingly. The hamiltonian $h_{0}$ acts in the Hilbert space of one
fermion. We define the following $\Gamma_{a}$ operators which act in the same
space:
\begin{equation}
\Gamma_{a}\equiv\left(  \Gamma^{\left(  \sigma\right)  },\Gamma_{\mu}^{\left(
\omega\right)  },\Gamma_{\mu a}^{\left(  \pi\right)  }\right)  =\left(
g_{\sigma}\beta,g_{\omega}\beta\gamma_{\mu},\frac{g_{A}}{2f_{\pi}}\beta
\tau_{a}\gamma_{5}\gamma_{\mu}\right)  \label{gamop}%
\end{equation}
as well as the operator:
\begin{equation}
\Theta_{a\left(  \mu b\right)  }^{\left(  \pi\right)  }=\frac{1}{4f_{\pi}^{2}%
}\beta\varepsilon_{abc}\tau_{c}\gamma_{\mu} \label{phiop}%
\end{equation}
The meson fields are denoted by:
\begin{equation}
S_{a}\left(  x\right)  \equiv\left(  \sigma\left(  x\right)  ,\left(
\partial_{\mu}\pi_{a}\left(  x\right)  \right)  ,\omega_{\mu}\left(  x\right)
\right)
\end{equation}
We will use the condensed notation:
\[
\left(  \Gamma S\right)  \equiv\Gamma^{\left(  \sigma\right)  }\sigma
+\Gamma_{\mu}^{\left(  \omega\right)  }\omega_{\mu}+\Gamma_{\mu a}^{\left(
\pi\right)  }\left(  \partial_{\mu}\pi_{a}\right)
\]%
\[
=g_{\sigma}\beta\sigma+\frac{g_{A}}{2f_{\pi}}\beta\tau_{a}\gamma_{5}%
\gamma_{\mu}\left(  \partial_{\mu}\pi_{a}\right)  +g_{\omega}\beta\gamma_{\mu
}\omega_{\mu}%
\]%
\begin{equation}
S\left(  \Theta S\right)  \equiv\pi_{a}\Theta_{a\left(  \mu b\right)  }\left(
\partial_{\mu}\pi_{b}\right)  =\frac{1}{4f_{\pi}^{2}}\beta\varepsilon
_{abc}\tau_{c}\pi_{a}\gamma_{\mu}\left(  \partial_{\mu}\pi_{b}\right)
\label{gsptp}%
\end{equation}
In the expressions above, $\gamma_{\mu}$ and $\omega_{\mu}$ are the euclidean
4-vectors defined in (\ref{omegamu}).\ Note that \emph{the coupling constants
are included in } $\Gamma$ and $\Theta$.

We further define the meson propagators $K$:
\begin{equation}
K=\left(
\begin{array}
[c]{ccc}%
\frac{1}{-\partial^{2}+m_{\sigma}^{2}} & 0 & 0\\
0 & \delta_{ab}\frac{1}{-\partial^{2}+m_{\pi}^{2}} & 0\\
0 & 0 & \frac{1}{\left(  -\partial^{2}+m_{\omega}^{2}\right)  }\left(
\delta_{\mu\nu}-\frac{1}{m_{\omega}^{2}}\partial_{\mu}\partial_{\nu}\right)
\end{array}
\right)  \label{kab}%
\end{equation}
With this condensed notation we can write:
\[
\frac{1}{2}SK^{-1}S=\int d^{4}x\left\{  \frac{1}{2}\sigma\left(  -\partial
^{2}+m_{\sigma}^{2}\right)  \sigma+\frac{1}{2}\pi_{a}\left(  -\partial
^{2}+m_{\pi}^{2}\right)  \pi_{a}\right.
\]%
\begin{equation}
\left.  +\frac{1}{2}\omega_{\mu}\left(  \left(  -\partial^{2}+m_{\omega}%
^{2}\right)  \delta_{\mu\nu}+\partial_{\mu}\partial_{\nu}\right)  \omega_{\nu
}\right\}  \label{kmin1}%
\end{equation}
and the euclidean actions (\ref{act}) and (\ref{act1}) acquire the form:%
\[
I\left(  N^{\dagger},N,S\right)  \equiv N^{\dagger}\left(  \partial_{\tau
}+h_{0}-\mu+\left(  \Gamma S\right)  +S\left(  \Theta S\right)  \right)
N+\frac{1}{2}SK^{-1}S
\]%
\begin{equation}
I\left(  S\right)  \equiv-Tr\ln\left(  \partial_{\tau}+h_{0}-\mu+\left(
\Gamma S\right)  +S\left(  \Theta S\right)  \right)  +\frac{1}{2}%
SK^{-1}S\label{actphi}%
\end{equation}
The partition function is given by:%
\[
Z=Tre^{-\beta\left(  H-\mu N\right)  }\equiv e^{W}=\int D\left(  N^{\dagger
},N\right)  D\left(  S\right)  e^{-N^{\dagger}\left(  \partial_{\tau}%
+h_{0}-\mu+\left(  \Gamma S\right)  +S\left(  \Theta S\right)  \right)
N-\frac{1}{2}SK^{-1}S}%
\]%

\begin{equation}
=\int D\left(  S\right)  e^{Tr\ln\left(  \partial_{\tau}+h_{0}-\mu+\left(
\Gamma S\right)  +S\left(  \Theta S\right)  \right)  -\frac{1}{2}SK^{-1}S}
\label{partfun}%
\end{equation}
The inclusion of separate chemical potentials for neutrons and protons would
cause no difficulty. In the zero temperature limit $\beta\rightarrow\infty$,
the energy $E$ of the system composed of $A$ nucleons is equal to:%
\begin{equation}
E-\mu A=-\frac{1}{\beta}W
\end{equation}

\section{One (boson) line irreducible diagrams.}
\label{sec:irrdiag}
\setcounter{equation}{0}\renewcommand{\theequation}{\arabic{section}%
.\arabic{equation}}

In this section we define the one (boson) line irreducible diagrams which are
used to construct the density functionals. The rules for calculating Feynman
diagrams are summarized in Appendices \ref{ap:diagclb} and \ref{ap:nucmesdiag}%
. Consider a general unlabeled connected diagram, such as:%
\begin{equation}%
{\parbox[b]{3.6789in}{\begin{center}
\includegraphics[
height=1.4633in,
width=3.6789in
]%
{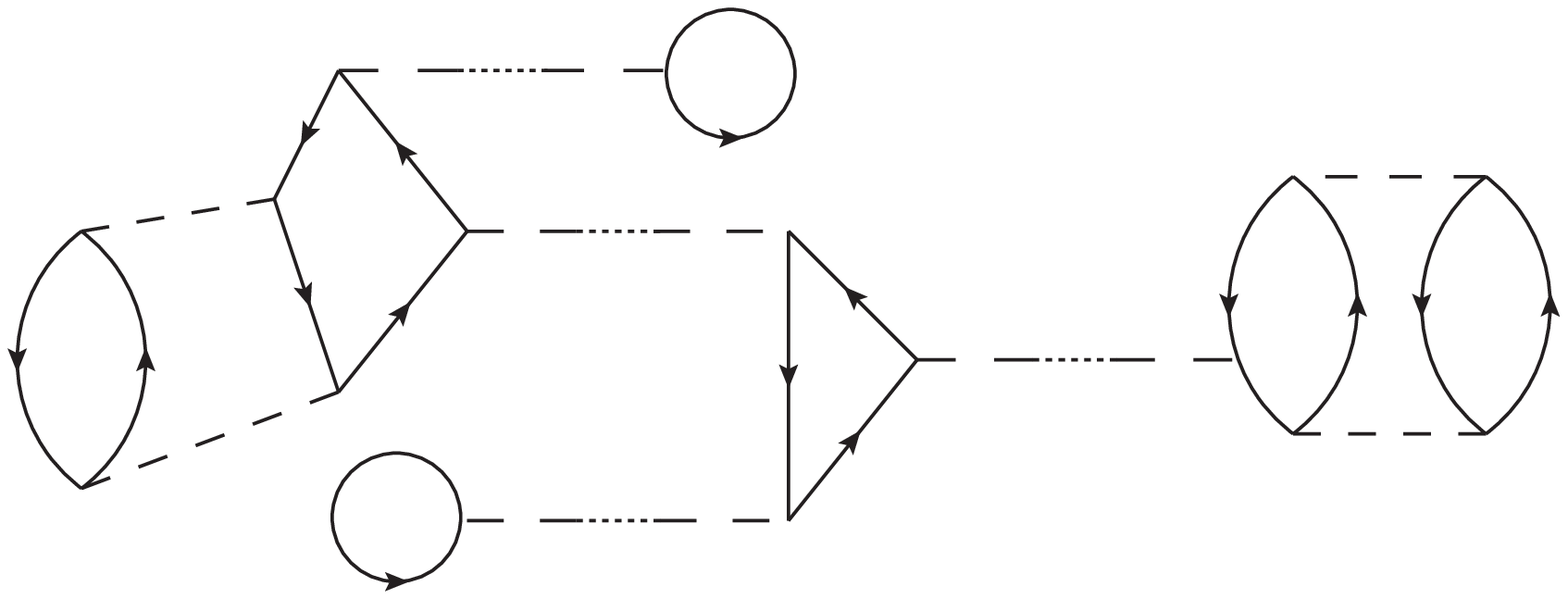}%
\\
{}%
\end{center}}}%
\label{genunlab}%
\end{equation}
We can recognize so-called \emph{articulation lines} which are dashed lines
(boson propagators or interaction lines) such that the diagram separates into
two disconnected parts when the dashed line is cut. In the diagram
(\ref{genunlab}) the four articulation lines are highlighted by dots. The
diagram (\ref{genunlab}) has $n_{a}=4$ articulation lines. When the
articulation lines are cut, the disconnected pieces consist of:

\begin{itemize}
\item \emph{cycles}, which are closed loops (formed by oriented fermion
propagators) the vertices of which are connected \emph{only} to articulation
lines. The diagram (\ref{genunlab}) has $n_{c}=3$ cycles.

\item \emph{one (boson) line irreducible parts}, which are irreducible in the
sense that they cannot be separated into two disconnected parts by cutting a
dashed line.\ The diagram (\ref{genunlab}) has $n_{I}=2$ irreducible parts.
\end{itemize}

Every diagram which has $n_{a}$ articulation lines, $n_{c}$ cycles and $n_{I}$
irreducible parts is such that:%
\begin{equation}
n_{I}+n_{c}-n_{a}=1 \label{nnn}%
\end{equation}
This topological property results from the fact that, whenever a boson
propagator is added to a diagram, it is either an articulation line or
not.\ If it is, it either adds a cycle or an irreducible part. If it is not,
it leaves $n_{I},n_{c}$ and $n_{a}$ unchanged.

The topological property (\ref{nnn}) does not hold for diagrams with open
ended meson dashed lines, which contain in fact a source factor $j$ at their
end point.\ The latter should be counted as an irreducible part and for such
diagrams the topological relation should read:%
\begin{equation}
n_{I}+n_{j}+n_{c}-n_{a}=1
\end{equation}
where $n_{j}$ is the number of open ends (or the number of source points) of
the diagram.

In the following we shall refer to one (boson) line irreducible diagrams
simply as one-line irreducible diagrams.

\section{The particle densities expressed in terms of one-line irreducible diagrams.}
\label{sec:selfconsist}
\setcounter{equation}{0}\renewcommand{\theequation}{\arabic{section}%
.\arabic{equation}}

In Appendices \ref{ap:denclb} and (\ref{ap:denmesnuc}) the particle densities
are expressed in terms of connected diagrams $\Gamma_{c}$.\ In the case of
fermions with Coulomb interactions, the particle density is given by
(\ref{partden}):%
\begin{equation}
n\left(  x\right)  =-tr\;\left\langle x\left|  g\right|  x\right\rangle +\int
d^{4}x_{1}d^{4}x_{2}\;\frac{\delta\Gamma_{c}}{\delta\left\langle x_{1}\left|
g\right|  x_{2}\right\rangle }\left\langle x_{1}\left|  g\right|
x\right\rangle \left\langle x\left|  g\right|  x_{2}\right\rangle
\label{partden2}%
\end{equation}
For nucleons interacting with mesons, the various densities are given by
(\ref{rhoax}):%
\begin{equation}
\rho_{a}\left(  x\right)  =-tr\;\left\langle x\left|  g\right|  x\right\rangle
\Gamma_{a}+\int d^{4}x_{1}d^{4}x_{2}\;\frac{\delta\Gamma_{c}}{\delta
\left\langle x_{1}\left|  g\right|  x_{2}\right\rangle }\left\langle
x_{1}\left|  g\right|  x\right\rangle \Gamma_{a}\left\langle x\left|
g\right|  x_{2}\right\rangle \label{rhoax2}%
\end{equation}
In (\ref{partden2}) and (\ref{rhoax2}), the connected diagrams are calculated
with the unperturbed fermion propagator:%
\begin{equation}
g=\frac{1}{\partial_{\tau}+h_{0}-\mu}\label{gusim}%
\end{equation}
and $\Gamma_{a}$ are the operators defined in (\ref{gamop}) and (\ref{phiop}%
).\footnote{No confusion should arise from the use of $\Gamma$ to denote
connected diagrams $\Gamma_{c}$ and the operators (\ref{gamop}).}

From here on we use the formalism applicable to nucleons interacting with
mesons. Is it however simple to recover expressions applicable to fermions
with Coulomb interactions.\ It suffices to set:%
\begin{equation}
\Gamma_{a}=1\;\;\;\;\;\;U_{a}\left(  x\right)  =U\left(  x\right)
\;\;\;\;\;\;\rho_{a}\left(  x\right)  =n\left(  x\right)  \label{clb}%
\end{equation}
In the case of nucleons interacting with mesons, the coupling constants are
included in the $\Gamma_{a}$. As a result, the particle densities $\rho
_{a}\left(  x\right)  $ are multiplied by the coupling constants and they are
related to the densities $n_{a}\left(  x\right)  $ in the usual sense by
equations (\ref{nax}). In the case of fermions with coulomb interactions, the
coupling constant $e^{2}$ is included in the interaction (\ref{coulpot}).

We now show that it is also possible to express the particle densities
$\rho_{a}\left(  x\right)  $ in terms of one-line irreducible diagrams
$\Phi\left(  U\right)  $ as follows:%
\[
\rho_{a}\left(  x\right)  =-\frac{\delta}{\delta U_{a}\left(  x\right)
}\left(  Tr\ln G^{-1}+\Phi\left(  U\right)  \right)
\]%
\begin{equation}
=-tr\;\left\langle x\left|  G\right|  x\right\rangle \Gamma_{a}+\int
d^{4}x_{1}d^{4}x_{2}\;\frac{\delta\Phi\left(  U\right)  }{\delta\left\langle
x_{1}\left|  G\right|  x_{2}\right\rangle }\left\langle x_{1}\left|  G\right|
x\right\rangle \Gamma_{a}\left\langle x\left|  G\right|  x_{2}\right\rangle
\label{rhoaxu}%
\end{equation}
In the expression (\ref{rhoaxu}), $G$ is the ''dressed'' fermion propagator:%
\begin{equation}
G=\frac{1}{\partial_{\tau}+h_{0}-\mu+U_{a}\Gamma_{a}} \label{bigg}%
\end{equation}
and $U_{a}\left(  x\right)  $ is a local potential which satisfies the
equation:%
\begin{equation}
U_{a}\left(  x\right)  =\int d^{4}y\;\left\langle x\left|  K_{ab}\right|
y\right\rangle \frac{\delta}{\delta U_{b}\left(  y\right)  }\;\left(  Tr\ln
G^{-1}+\Phi\left(  U\right)  \right)  \label{upotu}%
\end{equation}
For nucleons interacting with mesons, the boson propagators $K_{ab}$ are
defined in (\ref{kab}).\ For fermions with Coulomb interactions, $K_{ab}\equiv
K$, where $K$ is (minus) the Coulomb potential (\ref{coulpot}). The oriented
lines of the one-line irreducible diagrams $\Phi\left(  U\right)  $ are the
dressed propagators $G$. When $h_{0}$ is time independent, as assumed in this
work, the potentials $U_{a}\left(  x\right)  $ and the densities $\rho
_{a}\left(  x\right)  $ are also time independent because $\left\langle
x\left|  K_{ab}\right|  y\right\rangle $ depends only on $x_{\mu}-y_{\mu}$.

In this formulation, an approximation to the potential $U_{a}\left(  x\right)
$ and the particle density $\rho_{a}\left(  x\right)  $ are obtained by
choosing a subset $\Phi\left(  U\right)  $ of one-line irreducible diagrams.
Any finite or infinite subset can be chosen. For example, $\Phi\left(
U\right)  $ could be limited to the following one-line irreducible diagrams:%
\begin{equation}%
{\includegraphics[
height=0.7489in,
width=4.0542in
]%
{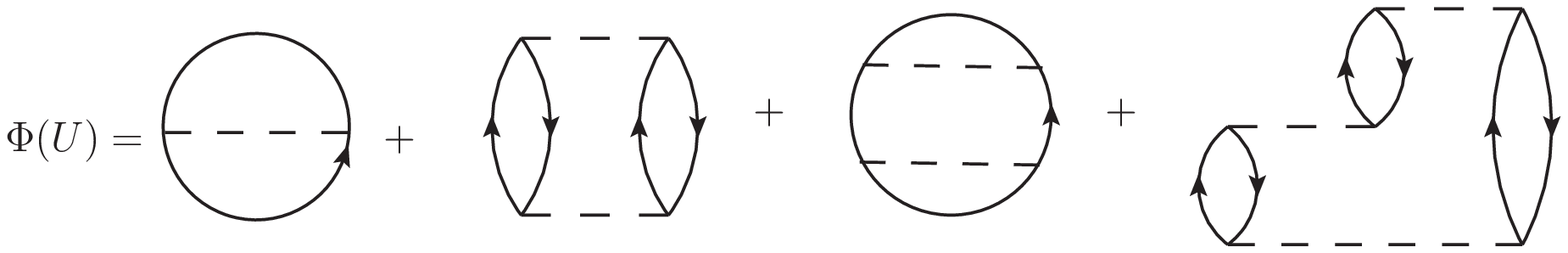}%
}%
\label{phiu}%
\end{equation}
When the set (\ref{phiu}) of one-line irreducible diagrams is chosen, the
density (\ref{rhoaxu}) is:%
\begin{equation}%
{\includegraphics[
height=1.6613in,
width=4.7288in
]%
{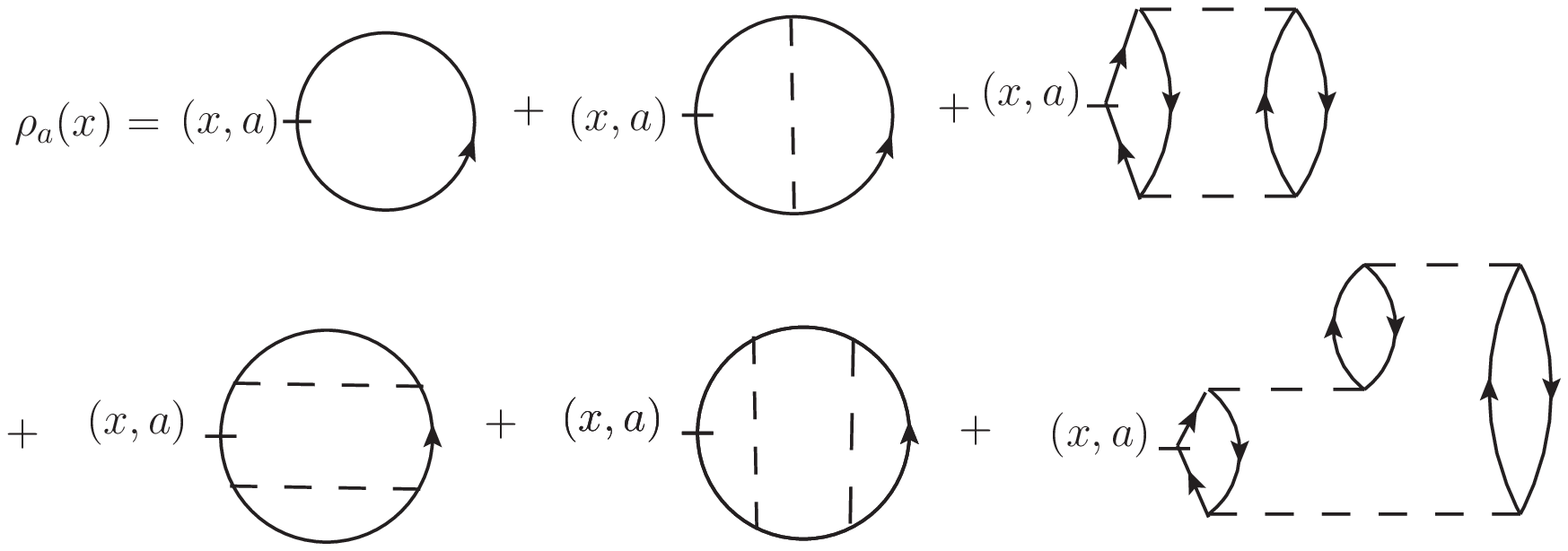}%
}%
\label{rhoaphiu}%
\end{equation}
The representation of the density in terms of diagrams with a slash labeled
$\left(  x,a\right)  $ is explained in Appendices \ref{ap:denclb} and
\ref{ap:denmesnuc}.

The potential (\ref{upotu}) can be written as follows:%
\begin{equation}
U_{a}\left(  x\right)  =\int d^{4}y\;\left\langle x\left|  K_{ab}\right|
y\right\rangle \left[  tr\;\left\langle y\left|  G\right|  y\right\rangle
\Gamma_{b}-\int d^{4}x_{1}d^{4}x_{2}\;tr\;\frac{\delta\Phi\left(  U\right)
}{\delta\left\langle x_{1}\left|  G\right|  x_{2}\right\rangle }\left\langle
x_{1}\left|  G\right|  y\right\rangle \Gamma_{b}\left\langle y\left|
G\right|  x_{2}\right\rangle \right]  \label{upotu2}%
\end{equation}
so that the diagram representation of the potential $U_{a}\left(  x\right)  $
is:%
\begin{equation}%
{\includegraphics[
height=1.4434in,
width=5.0246in
]%
{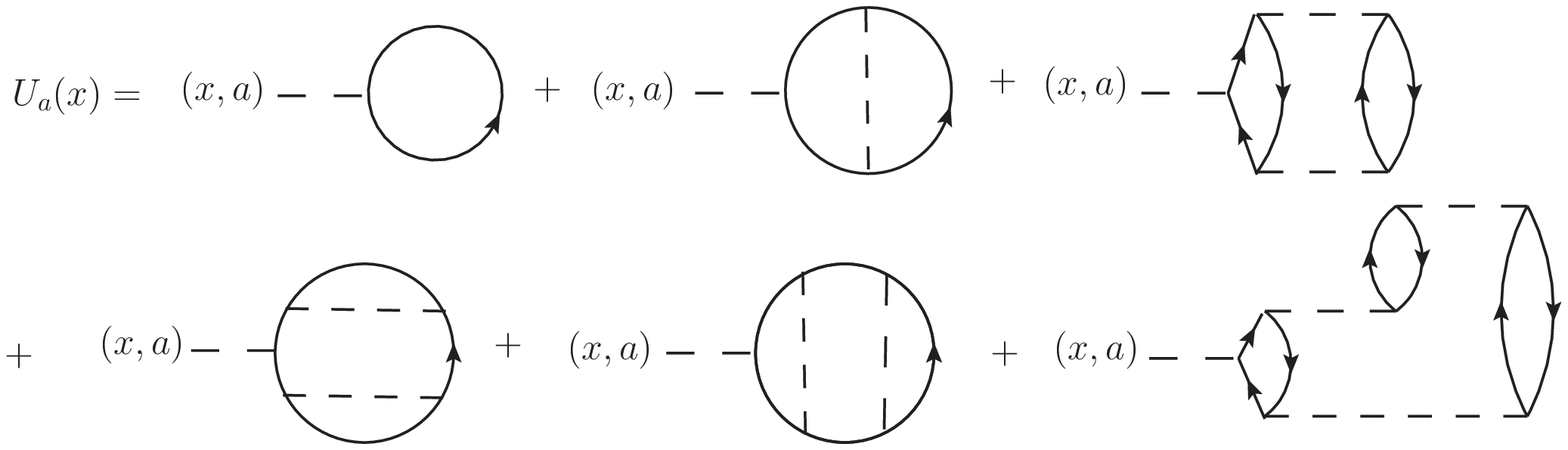}%
}%
\label{uphiu}%
\end{equation}
In the diagrams (\ref{rhoaphiu}) and (\ref{uphiu}) the oriented lines are the
dressed fermion propagators (\ref{bigg}).

The dressed propagator $G$ can be expressed in terms of the unperturbed
propagator $g$ by the series:%
\begin{equation}
G=g-gU\Gamma g+gU\Gamma gU\Gamma g-...
\end{equation}
Therefore each oriented fermion propagator $G$ in the diagrams (\ref{rhoaphiu}%
) and (\ref{uphiu}) generates an infinite set of diagrams composed of oriented
lines $g$ with any number of $U$ insertions. Furthermore, iterations of
equation (\ref{upotu2}) will generate, for each one of these insertions, an
infinite set of diagrams connected by an articulation line. It is easy to
check that the diagrams (\ref{uphiu}) generate diagrams with a tree structure
and that they yield a particle density (\ref{rhoax2}) and a potential
(\ref{upotu}) in which the contribution of each diagram $\Gamma_{c}$ is
included once and only once.

In many cases, the ground state densities may vanish because of
self-consistent symmetries, discussed in Ref.\cite{Ripka86}, such as spherical
symmetry, time-reversal symmetry, isospin or flavor symmetry for example.
Self-consistent symmetries depend on the system which is being calculated,
whether, for example, it is in a rotating frame, whether it is exposed to an
external field, whether the number of nucleons is even or odd, whether there
is an excess of neutrons or protons, and so forth. In this work we do not
assume such symmetries.

Note that the potentials $U_{a}\left(  x\right)  $ and the densities $\rho
_{a}\left(  x\right)  $ are related the simple analytic expressions:%
\begin{equation}
U_{a}\left(  x\right)  =-\int d^{4}y\;\left\langle x\left|  K_{ab}\right|
y\right\rangle \rho_{b}\left(  y\right)  \;\;\;\;\;\;\;\;\rho_{a}\left(
x\right)  =-\int d^{4}y\;\left\langle x\left|  K_{ab}^{-1}\right|
y\right\rangle U_{b}\left(  y\right)  \label{uaxkrhoy}%
\end{equation}
This is in sharp contrast with the potentials which are expressed in terms of
Kohn-Sham orbits (see Section \ref{sec:kohnsham}).

In Section \ref{sec:partirr} we express the partition function in terms of
one-line irreducible diagrams and in Section \ref{sec:stat} we show that it is
a stationary functional of both the particle densities $\rho$ and the
potentials $U$.

\section{Iteration procedure to calculate the particle densities and potentials.}
\label{sec:iterate}
\setcounter{equation}{0}\renewcommand{\theequation}{\arabic{section}%
.\arabic{equation}}

The time-independent single particle hamiltonian $h_{0}+U_{a}\Gamma_{a}$ can
be diagonalized:%
\begin{equation}
\left(  h_{0}+U_{a}\Gamma_{a}\right)  \left|  \lambda\right\rangle
=e_{\lambda}\left|  \lambda\right\rangle \label{fermorbs}%
\end{equation}
The eigenstates $\left|  \lambda\right\rangle $ are fermion orbits and we
distinguish particle orbits $\left|  p\right\rangle $ which have energies
$e_{p}>\mu$ and hole orbits $\left|  h\right\rangle $ with energies $e_{h}%
<\mu$. The eigenvalue problem (\ref{fermorbs}) involves the local and static
potentials $U_{a}\left(  x\right)  $.\ This makes the computation of the
fermion orbits $\left|  \lambda\right\rangle $ considerably simpler and faster
than, for example, the determination of the poles and residues of the 
single particle Green's function, which involves a non-local and energy 
dependent mass operator. This is the
main interest of constructing density functionals.

In the zero temperature limit, the dressed propagator (\ref{bigg}) can be
expressed in terms of the particle and hole orbits as in (\ref{g12}) and
(\ref{g12zerot}):%
\[
\left\langle x_{1}\left|  G\right|  x_{2}\right\rangle \equiv\left\langle
\tau_{1}\vec{r}_{1}\left|  G\right|  \tau_{2}\vec{r}_{2}\right\rangle
\]%
\[
=\theta\left(  \tau_{1}-\tau_{2}\right)  \sum_{p}e^{-\left(  e_{p}-\mu\right)
\left(  \tau_{1}-\tau_{2}\right)  }\left\langle \left.  \vec{r}_{1}\right|
p\right\rangle \left\langle \left.  p\right|  \vec{r}_{2}\right\rangle
-\theta\left(  \tau_{2}-\tau_{1}\right)  \sum_{h}e^{-\left(  e_{h}-\mu\right)
\left(  \tau_{1}-\tau_{2}\right)  }\left\langle \left.  \vec{r}_{1}\right|
h\right\rangle \left\langle \left.  h\right|  \vec{r}_{2}\right\rangle
\]%
\begin{equation}
\left\langle \tau\vec{r}_{1}\left|  G\right|  \tau\vec{r}_{2}\right\rangle
=-\sum_{h}\left\langle \left.  \vec{r}_{1}\right|  h\right\rangle \left\langle
\left.  h\right|  \vec{r}_{2}\right\rangle \label{g12bis}%
\end{equation}
In this representation, the first term of (\ref{rhoaxu}) (but not the second)
acquires the form of a density of an uncorrelated Slater determinant:%
\begin{equation}
-tr\;\left\langle x\left|  G\right|  x\right\rangle \Gamma_{a}=\sum
_{h}\left\langle \left.  h\right|  \vec{r}\right\rangle \Gamma_{a}\left\langle
\left.  \vec{r}\right|  h\right\rangle
\end{equation}

The equations (\ref{upotu}) or (\ref{upotu2}) can be solved by iteration.\ We
begin by choosing a set $\Phi\left(  U\right)  $ of one-line irreducible
diagrams. Then:

\begin{enumerate}
\item  We make an initial guess at the potentials $U_{a}\left(  x\right)
$.\ Alternatively, we can make an initial guess at the particle densities
$\rho_{a}\left(  x\right)  $ and deduce the initial potentials from
(\ref{uaxkrhoy}).

\item  With the potentials $U_{a}\left(  x\right)  $ we calculate the fermion
orbits (\ref{fermorbs}) by diagonalizing $h_{0}+U_{a}\Gamma_{a}$. This yields
the dressed fermion propagator $G$ defined in (\ref{g12bis}).

\item  We express the set $\Phi\left(  U\right)  $ of one-line irreducible
diagrams in terms of the fermion propagator $G$ and we recalculate the
potentials $U_{a}\left(  x\right)  $ using the equations (\ref{upotu}) or
(\ref{upotu2}).

\item  We return to step 2 and we continue the process until convergence is achieved.
\end{enumerate}

The densities $\rho_{a}\left(  x\right)  $ can then be deduced from
(\ref{uaxkrhoy}). The equation (\ref{rhoax2}) can be used to replace the
iteration procedure above by one which yields successive approximations to the densities.

The fact that it is equally easy to calculate either the densities or the
potentials is due to the simple equation (\ref{uaxkrhoy}) which relates the
two.\ This is why we call the eigenstates (\ref{fermorbs}) $h_{0}+U_{a}%
\Gamma_{a}$ the \emph{optimal orbits} for constructing stationary density
functionals. In Section \ref{sec:kohnsham} we show that the use of Kohn-Sham
orbits is more complicated.

\section{The partition function expressed in terms of one-line irreducible diagrams.}

\label{sec:partirr}

In this section we express the partition function in terms of one-line
irreducible diagrams.\ The partition function acquires the form of a
stationary functional, namely (\ref{wu}), of either the potentials
$U_{a}\left(  x\right)  $ or the particle densities $\rho_{a}\left(  x\right)
$.

In Appendices \ref{ap:diagclb} and \ref{ap:nucmesdiag} the partition function
$Z=e^{W}$ is expressed as a sum of connected diagrams $\Gamma_{c}$, calculated
with unperturbed fermion propagators $g$, defined in (\ref{gusim}):%
\begin{equation}
W=Tr\ln g^{-1}+\Gamma_{c}\label{wg}%
\end{equation}
If we wish to express the partition function in terms of one-line irreducible
diagrams $\Phi\left(  U\right)  $, calculated with the dressed propagators $G$
defined in (\ref{bigg}), we need to correct for the fact that the one-line
irreducible diagrams $\Phi\left(  U\right)  $ overcount the diagrams
$\Gamma_{c}$. Indeed, each diagram $\Gamma_{c}$, which can be decomposed into
$n_{I}\left(  \Gamma_{c}\right)  $ one-line irreducible parts, is included
$n_{I}\left(  \Gamma_{c}\right)  $ times in the corresponding one-line
irreducible diagram $\Phi\left(  U\right)  $. Consider, for example, the
one-line irreducible diagram:
\begin{equation}%
{\includegraphics[
height=0.7965in,
width=1.407in
]%
{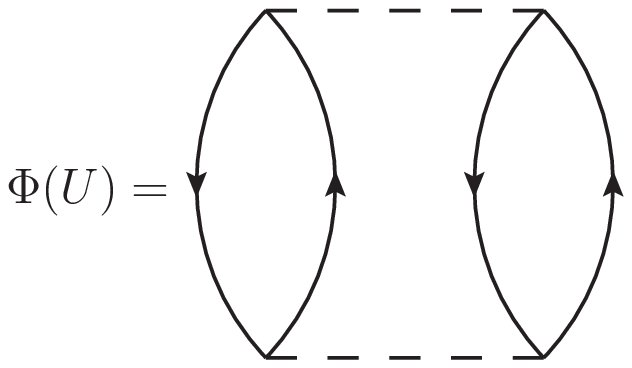}%
}%
\label{twopart}%
\end{equation}
which is calculated with dressed propagators $G$.\ The connected diagram:
\begin{equation}%
{\includegraphics[
height=0.6962in,
width=2.1145in
]%
{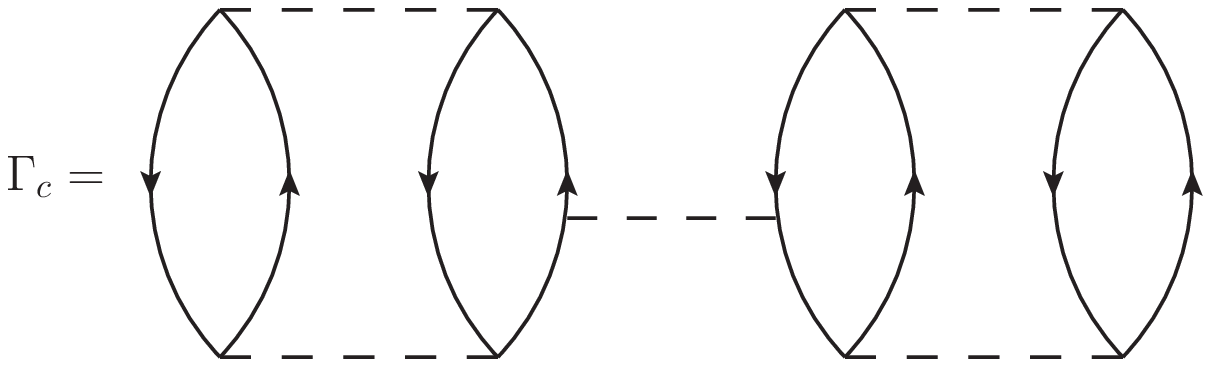}%
}%
\label{twopart2}%
\end{equation}
which is calculated with bare propagators $g$ and which contains $n_{I}=2$
one-line irreducible parts, is included twice in the diagram $\Phi\left(
U\right)  $. Indeed either of the two irreducible parts of the diagram
(\ref{twopart2}) can be considered as dressing the propagator of the other.
More generally, a set $\Phi\left(  U\right)  $ of one-line irreducible
diagrams, calculated dressed propagators $G$, generates a set of connected
diagrams $\Gamma_{c}$ composed of $n_{I}\left(  \Gamma_{c}\right)  $
irreducible parts and it overcounts them $n_{I}\left(  \Gamma_{c}\right)  $
times.\ Its contribution is therefore equal to:
\begin{equation}
\Phi\left(  U\right)  =\sum_{\Gamma_{c}}n_{I}\left(  \Gamma_{c}\right)
\;\Gamma_{c}\label{phiuc}%
\end{equation}

There are two other ways to generate connected diagrams. The connected
diagrams $\Gamma_{c}$ generated by the one-line irreducible diagrams
$\Phi\left(  U\right)  $ can also be generated by the following set of
diagrams, consisting of a fermion closed loop interacting once, twice,... with
the potential $U_{a}\Gamma_{a}$:
\begin{equation}%
{\includegraphics[
height=1.1165in,
width=3.7628in
]%
{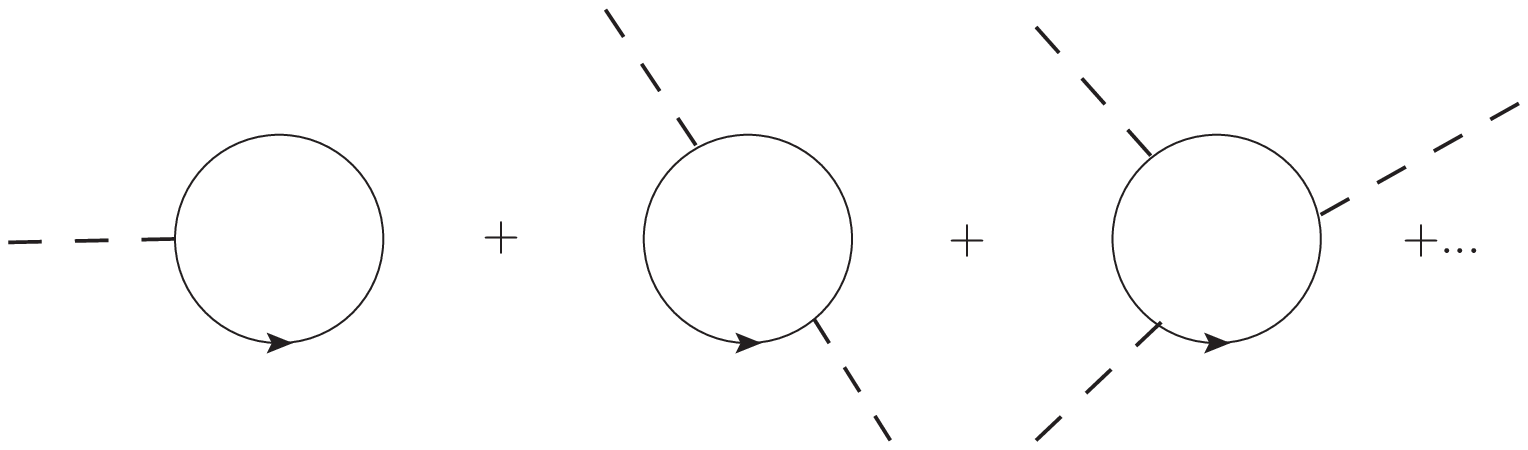}%
}%
\label{cycle}%
\end{equation}
In these diagrams the boson lines appended to the closed oriented fermion line
are interactions with the potential $U_{a}\Gamma_{a}$. The contribution of the
diagrams (\ref{cycle}) is:
\[
tr\int d^{4}x\;\left\langle x\left|  g\Gamma_{a}U_{a}-\frac{1}{2}g\Gamma
_{a}U_{a}g\Gamma_{b}U_{b}+\frac{1}{3}g\Gamma_{a}U_{a}g\Gamma_{b}U_{b}%
g\Gamma_{c}U_{c}-...\right|  x\right\rangle
\]%
\begin{equation}
=Tr\ln\left(  1+g\Gamma_{a}U_{a}\right)  =Tr\ln gG^{-1}\label{loopc}%
\end{equation}
where we used the fact that $gG^{-1}=1+g\Gamma_{a}U_{a}$. However, not only
does the expression (\ref{loopc}) include all the diagrams generated by the
one-line irreducible diagrams $\Phi\left(  U\right)  $, but it overcounts
them. Indeed, each diagram $\Gamma_{c}$, which can be decomposed into
$n_{c}\left(  \Gamma_{c}\right)  $ fermion loops, occurs $n_{c}\left(
\Gamma_{c}\right)  $ times in the contribution (\ref{loopc}), which is
therefore equal to:
\begin{equation}
Tr\ln gG^{-1}=\sum_{\Gamma_{c}}n_{c}\left(  \Gamma_{c}\right)  \;\Gamma
_{c}\label{nccon}%
\end{equation}
A third way to include the diagrams generated by the set $\Phi\left(
U\right)  $ is to calculate the expression:
\[%
\raisebox{-0cm}{\includegraphics[
height=0.5293in,
width=1.1355in
]%
{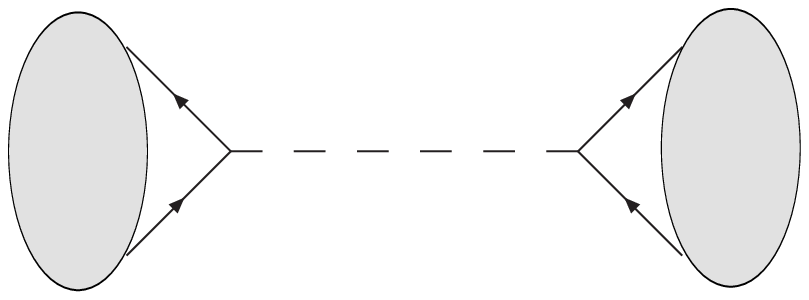}%
}%
\]%
\begin{equation}
=\frac{1}{2}\int d^{4}x\int d^{4}y\;\rho_{a}\left(  x\right)  \left\langle
x\left|  K_{ab}\right|  y\right\rangle \rho_{b}\left(  y\right)  =\frac{1}%
{2}\int d^{4}x\int d^{4}y\;U_{a}\left(  x\right)  \left\langle x\left|
K_{ab}^{-1}\right|  y\right\rangle U_{b}\left(  y\right)  \label{rhokrho}%
\end{equation}
where $\rho_{a}\left(  x\right)  $ is the density (\ref{rhoaxu}). Not only
does the expression (\ref{rhokrho}) include all the diagrams generated by the
one-line irreducible diagrams $\Phi\left(  U\right)  $, but each diagram
$\Gamma_{c}$ which contains $n_{a}\left(  \Gamma_{c}\right)  $ articulation
lines, is included $n_{a}\left(  \Gamma_{c}\right)  $ times in the
contribution (\ref{rhokrho}), so that:
\begin{equation}
\frac{1}{2}\int d^{4}x\int d^{4}y\;U_{a}\left(  x\right)  \left\langle
x\left|  K_{ab}^{-1}\right|  y\right\rangle U_{b}\left(  y\right)
=\sum_{\Gamma_{c}}n_{a}\left(  \Gamma_{c}\right)  \;\Gamma_{c}\label{nacon}%
\end{equation}
We can combine the results (\ref{phiuc}), (\ref{nccon}) and (\ref{nacon}) to
write:
\[
\Phi\left(  U\right)  +Tr\ln gG^{-1}-\frac{1}{2}\int d^{4}x\int d^{4}%
y\;U_{a}\left(  x\right)  \left\langle x\left|  K_{ab}^{-1}\right|
y\right\rangle U_{b}\left(  y\right)
\]%
\begin{equation}
=\sum_{\Gamma_{c}}\left[  n_{I}\left(  \Gamma_{c}\right)  +n_{c}\left(
\Gamma_{c}\right)  -n_{a}\left(  \Gamma_{c}\right)  \right]  \;\Gamma_{c}%
=\sum_{\Gamma_{c}}\Gamma_{c}\label{sumg}%
\end{equation}
where we used the topological relation (\ref{nnn}). The expression
(\ref{sumg}) therefore sums the diagrams $\Gamma_{c}$ generated by the set
$\Phi\left(  U\right)  $ of one-line irreducible diagrams correctly (meaning
that each one is counted once).

The partition function $Z=e^{W}$ can therefore be expressed in terms of
one-line irreducible diagrams as follows:
\begin{equation}
\ln Z=W=tr\ln g^{-1}+\Phi\left(  U\right)  +Tr\ln gG^{-1}-\frac{1}{2}UK^{-1}U
\end{equation}
which reduces to:%
\begin{equation}
W=Tr\ln G^{-1}+\Phi\left(  U\right)  -\frac{1}{2}UK^{-1}U\label{wu}%
\end{equation}
In the zero temperature limit, the energy $E$ of the system composed of $N$
particles is related to $W$ by the equation:%
\begin{equation}
W=-\beta\left(  E-\mu N\right)
\end{equation}

\section{Stationary properties of the density functional.}

\label{sec:stat}

Consider a variation $U_{a}\rightarrow U_{a}+\delta U_{a}$ of the potential
$U_{a}$. The corresponding variation of $W$ is:
\begin{equation}
\delta W=\int d^{4}x\;\left[  \delta U_{a}\left(  x\right)  \frac{\delta
}{\delta U_{a}\left(  x\right)  }\left[  Tr\ln G^{-1}+\Phi\left(  U\right)
\right]  -\delta U_{a}\left(  x\right)  \int d^{4}y\left\langle x\left|
K_{ab}^{-1}\right|  y\right\rangle U_{b}\left(  y\right)  \right]
\end{equation}
The equation $\frac{\delta W}{\delta U_{a}\left(  x\right)  }=0$ therefore
leads to the equation (\ref{upotu}). In other words, the equation
(\ref{upotu}) which determines the potential $U_{a}\left(  x\right)  $ is
equivalent to the equation which states that the functional (\ref{wu}) should
be stationary with respect to variations of the potential $U_{a}\left(
x\right)  $.

Since $U=-K\rho$, the generating functional (\ref{wu}) may equally well be
regarded a functional of the densities $\rho_{a}$. And since $\frac{\delta
W}{\delta\rho}=\frac{\delta W}{\delta U}\frac{\delta U}{\delta\rho}=-K$
$\frac{\delta W}{\delta U}$, the functional (\ref{wu}) is also a stationary
functional of the particle densities $\rho_{a}$.

If all one-line irreducible diagrams $\Phi$ are neglected, the scheme above
reduces to the Hartree approximation. If all one-line irreducible diagrams are
included (a goal never achieved) the scheme yields an exact energy and exact
particle densities. An approximation is defined in terms of a selected subset
of one-line irreducible diagrams. In all cases, the fermion orbits are
eigenstates of a single-particle hamiltonian involving a local potential.

\section{One-line irreducible diagrams identified in the path integral.}
\setcounter{equation}{0}\renewcommand{\theequation}{\arabic{section}%
.\arabic{equation}}

\label{sec:pathint}

The expression (\ref{wu}) has been derived from an analysis of the topology of
connected diagrams.\ It can also be derived by an algebraic manipulation of the
path integral (\ref{partfun}) of the partition function\cite{Ripka86}%
.\ Indeed, let us make the following change of the integration variable:
\begin{equation}
S_{a}\left(  x\right)  \rightarrow S_{a}^{\prime}\left(  x\right)
=S_{a}\left(  x\right)  +U_{a}\left(  x\right)
\end{equation}
where $U_{a}\left(  x\right)  $ is an as yet unspecified local field. The
partition function (\ref{partfun}) becomes, after dropping the primes:
\begin{equation}
e^{W}=\int D\left(  S\right)  e^{Tr\ln\left(  \partial_{\tau}+h_{0}-\mu+\Gamma
S+\Gamma U+S\left(  \Phi S\right)  +U\left(  \Phi U\right)  +U\left(  \Phi
S\right)  +S\left(  \Phi U\right)  \right)  -\int d^{4}x\;\left(  \frac{1}%
{2}SK^{-1}S+SK^{-1}U+\frac{1}{2}UK^{-1}U\right)  }%
\end{equation}
Let us define:
\begin{equation}
G^{-1}=\partial_{\tau}+h_{0}-\mu+\Gamma U+U\left(  \Phi U\right)
\end{equation}
so that:%
\[
Tr\ln\left(  \partial_{\tau}+h_{0}-\mu+\Gamma S+\Gamma U+S\left(  \Phi
S\right)  +U\left(  \Phi U\right)  +U\left(  \Phi S\right)  +S\left(  \Phi
U\right)  \right)
\]%
\begin{equation}
=Tr\ln G^{-1}+tr\ln\left(  1+G\Gamma S+GU\left(  \Phi S\right)  +GS\left(
\Phi S\right)  \right)
\end{equation}
The partition function becomes:
\begin{equation}
e^{W}=e^{-Tr\ln G-\frac{1}{2}UK^{-1}U}\int D\left(  S\right)  e^{Tr\ln\left(
1+G\Gamma S+GU\left(  \Phi S\right)  +GS\left(  \Phi S\right)  \right)  -\int
d^{4}x\;\left(  \frac{1}{2}SK^{-1}S+SK^{-1}U\right)  }%
\end{equation}
so that:\qquad%
\begin{equation}
W=Tr\ln G^{-1}-\frac{1}{2}UK^{-1}U
\end{equation}%
\begin{equation}
+\ln\int D\left(  S\right)  e^{Tr\ln\left(  1+G\Gamma S+GU\left(  \Phi
S\right)  +GS\left(  \Phi S\right)  \right)  -\int d^{4}x\;\left(  \frac{1}%
{2}SK^{-1}S+SK^{-1}U\right)  }%
\end{equation}
If we compare this to the expression (\ref{wu}), we obtain a path
integral representation of one-line irreducible diagrams:%

\begin{equation}
\Phi\left(  U\right)  =\ln\int D\left(  S\right)  e^{Tr\ln\left(  1+G\Gamma
S+GU\left(  \Phi S\right)  +GS\left(  \Phi S\right)  \right)  -\int
d^{4}x\;\left(  \frac{1}{2}SK^{-1}S+SK^{-1}U\right)  } \label{ineir}%
\end{equation}
To check that the path integral generates only one-line irreducible diagrams,
we expand $Tr\ln\left(  1+G\Gamma S+GU\left(  \Phi S\right)  +GS\left(  \Phi
S\right)  \right)  $ in powers of $\left(  \Gamma S+U\left(  \Phi S\right)
+S\left(  \Phi S\right)  \right)  $ and we derive the corresponding Feynman
diagrams as done in Appendices \ref{ap:diagclb} and \ref{ap:nucmesdiag}.
However, the difference with
the diagram expansion of the expression (\ref{partfun}) is due to the
occurrence of the term $SK^{-1}U$ in (\ref{ineir}), which generates diagrams
with open ended interaction lines which are attached to fermion propagators at
one end and to the potential $U$ at the other.\ These diagrams are one-line
reducible. The reader can check for himself that, when the potential $U$ is
chosen to be $U=-K\rho$, these diagrams cancel the one-line reducible diagrams
(without open ended interaction lines) so that the expression (\ref{ineir})
generates only one-line irreducible diagrams.

\section{The use of Kohn-Sham orbits.}

\label{sec:kohnsham}

We now compare the theory expressed in Sections \ref{sec:selfconsist} and
\ref{sec:partirr} to the construction of stationary density functionals
using Legendre transforms.\ For simplicity, we limit the discussion to systems
of fermions with Coulomb interactions, which involve only one single-particle
density $n\left(  x\right)  $ and for which we can set:%
\begin{equation}
\Gamma_{a}=1\;\;\;\;\;\;U_{a}\left(  x\right)  =U\left(  x\right)
\;\;\;\;\;\;\rho_{a}\left(  x\right)  =n\left(  x\right)
\end{equation}
Let us add to $h_{0}$ a source term $J\left(  x\right)  $ coupled to the
particle density $n\left(  x\right)  $.\ A stationary density functional is
then obtained from the Legendre transform:%
\begin{equation}
\Gamma\left(  n\right)  =W+Jn
\end{equation}
The Lagrange multiplier $J\left(  x\right)  $ acts as a local potential. The
equation $\frac{\delta\Gamma}{\delta n}=J$ states that, for a given $J$, the
density is given by $n=-\frac{\delta W}{\delta J}$ and the equation
$\frac{\delta\Gamma}{\delta n}=0$ states that $J=0$, in which case the density
$n$ is the equilibrium or ground state density. The inversion of the equation
$n=-\frac{\delta W}{\delta J}$ consists in finding the local potential
$J\left(  x\right)  $ which yields a given density $n\left(  x\right)  $,
thereby making $\Gamma\left(  n\right)  $ a stationary functional of the
density. A frequently used way to invert the equation $n=-\frac{\delta
W}{\delta J}$ consists in separating the source term $J$ into two parts:%
\begin{equation}
J=J_{0}+J_{int}%
\end{equation}
where $J_{0}$ is chosen such that the exact density $n\left(  x\right)  $ is
given by the uncorrelated Slater determinant, the orbits of which are
eigenstates of the hamiltonian $h_{0}+J_{0}$ and which are called Kohn-Sham
orbits (see \cite{Fukuda1994},\cite{Fukuda1995}, \cite{Okumura1996}%
,\cite{ValievFernando1998},\cite{Polonyi2004},\cite{Furnstahl2009-1} and
references therein). The ''interacting'' part $J_{int}$ is then calculated in
terms of the density $n$ and the potential $J_{0}$ by considering, for
example, an expansion in powers of the coupling constant, expressed by
connected Feynman or Goldstone diagrams, calculated with fermion propagators
$G=\frac{1}{\partial_{\tau}+h_{0}-\mu+J_{0}}$ which can be constructed from
the Kohn-Sham orbits. The interacting part $J_{int}$ can then be expressed in
terms of one-line irreducible diagrams with, in addition, extra diagrams
involving for example the inverse density-density 
correlation function (\ref{drrm1}).
Finally, when $J_{int}$ is determined, $J_{0}$ is determined by the equation
$J_{0}=-J_{int}$ which states that, in the ground state, $J=0$.

In section \ref{sec:selfconsist} we expressed the density directly in terms of
one-line irreducible diagrams, which are calculated with a self-consistent
potential $U\left(  x\right)  $, determined by the equation (\ref{upotu}). The
potential $U\left(  x\right)  $ is then simply and analytically related to the
density $n\left(  x\right)  $ by the equation (\ref{uaxkrhoy}).
We now show that if we split the self-consistent potential $U\left(  x\right)
$ into two terms:%
\begin{equation}
U\left(  x\right)  =U_{0}\left(  x\right)  +U_{int}\left(  x\right)
\label{u0uint}%
\end{equation}
so that the dressed fermion propagator becomes:%
\begin{equation}
G^{-1}=G_{0}^{-1}+U_{int}\;\;\;\;\;\;\;\;G=G_{0}-G_{0}U_{int}G\label{goint}%
\end{equation}
with:%
\begin{equation}
G_{0}=\frac{1}{\partial_{\tau}+h_{0}-\mu+U_{0}}\label{gu0}%
\end{equation}
and if we choose the potential $U_{0}\left(  x\right)  $ such that the density
$n\left(  x\right)  $ is equal to:%
\begin{equation}
n\left(  x\right)  =-tr\;\left\langle x\left|  G_{0}\right|  x\right\rangle
\label{ksrho}%
\end{equation}
then $U_{0}$ becomes identical to the source term $J_{0}$ used in the Legendre
transform method.

There is no real need to separate the potential into the two terms
(\ref{u0uint}). In fact we shall see that it makes the theory and the
calculations somewhat more complicated.\ We do so however, to compare
the present theory to the frequently used Kohn-Sham orbits in 
the Legendre transform method.

We can express equation (\ref{ksrho}) in terms of the 
eigenstates of $h_{0}+U_{0}$:%
\begin{equation}
\left(  h_{0}+U_{0}\right)  \left|  \lambda\right\rangle =e_{\lambda}\left|
\lambda\right\rangle \label{ksorbs}%
\end{equation}
and ''particle'' and ''hole'' orbits $\left|  p\right\rangle $ and $\left|
h\right\rangle $ are the eigenstates $\left|  \lambda\right\rangle $ belonging
to eigenvalues respectively $e_{p}>\mu$ and $e_{h}<\mu$. Since $G_{0}$ can be
expressed in the form (\ref{g12}) and (\ref{g12zerot}), the density
(\ref{ksrho}) can be written in the form:%

\[
n\left(  x\right)  =-tr\;\left\langle x\left|  G_{0}\right|  x\right\rangle
=\sum_{h}\left\langle \left.  h\right|  \vec{r}\right\rangle \left\langle
\left.  \vec{r}\right|  h\right\rangle
\]%
\begin{equation}%
\raisebox{-0cm}{\includegraphics[
height=0.4687in,
width=0.7325in
]%
{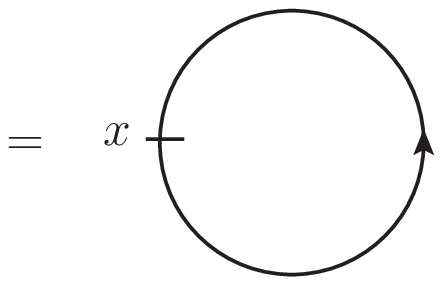}%
}%
\label{densham}%
\end{equation}
The equations (\ref{ksrho}) and (\ref{densham}) state the density of the
correlated system is equal to the density of a Slater determinant composed of
the eigenstates of $h_{0}+U_{0}$.\ That is why the eigenstates (\ref{ksorbs})
are usually called Kohn-Sham orbits.

\subsection{Equations for the potentials $U_{0}$ and $U_{int}$.}

The equation (\ref{rhoaxu}) which expresses the density $n\left(  x\right)  $
in terms of one-line irreducible diagrams can be written in the form:%
\begin{equation}
n\left(  x\right)  =-\;tr\;\left\langle x\left|  G_{0}\right|  x\right\rangle
-\;tr\;\left\langle x\left|  G_{0}U_{int}G\right|  x\right\rangle
-\frac{\delta\Phi\left(  U\right)  }{\delta U\left(  x\right)  }%
\end{equation}
where (\ref{goint}) was used. With the choice (\ref{ksrho}) of $U_{0}$ this
equation reduces to:%
\begin{equation}
tr\;\left\langle x\left|  G_{0}U_{int}G\right|  x\right\rangle =\left.
\frac{\delta\Phi\left(  U\right)  }{\delta U\left(  x\right)  }\right|
_{U=U_{0}+U_{int}} \label{ueq2}%
\end{equation}
This is an integral equation for $U_{int}$ which can be solved by iteration,
given the potential $U_{0}$.\ Once $U_{int}$ is thus determined, $U_{0}$ can
be obtained from the equation (\ref{uaxkrhoy}), which reads:%
\begin{equation}
U_{0}\left(  x\right)  +U_{int}\left(  x\right)  =-\int d^{4}x^{\prime
}\;\left\langle x\left|  K\right|  x^{\prime}\right\rangle n\left(  x^{\prime
}\right)  \label{u12rho}%
\end{equation}
Since $K$ is (minus) the Coulomb potential (\ref{coulpot}), we obtain:%
\begin{equation}
U_{0}\left(  \vec{r}\right)  =\int d^{3}r^{\prime}\;\frac{e^{2}}{4\pi\left|
\vec{r}-\vec{r}^{\prime}\right|  }n\left(  \vec{r}^{\prime}\right)
-U_{int}\left(  \vec{r}\right)  \label{u0fromu1}%
\end{equation}
When $U=U_{0}+U_{int}$ the functional (\ref{wu}) becomes:%
\begin{equation}
W\left(  U\right)  =Tr\ln G_{0}^{-1}+Tr\ln\left(  1+G_{0}U_{int}\right)
+\Phi\left(  U_{0}+U_{int}\right)  -\frac{1}{2}\left(  U_{0}+U_{int}\right)
K^{-1}\left(  U_{0}+U_{int}\right)  \label{wu2}%
\end{equation}
Once $U_{0}$ and $U_{int}$ have been determined, we can use (\ref{u12rho}) to
simplify $W\left(  U\right)  $ to:%
\begin{equation}
W =Tr\ln G_{0}^{-1}-\frac{1}{2}nKn+Tr\ln\left(
1+G_{0}U_{int}\right)  +\Phi\left(  U_{0}+U_{int}\right)  \label{wuint}%
\end{equation}

\subsection{First order approximation.}

Let us first consider the case where the one-line irreducible diagrams
$\Phi\left(  U\right)  $ are limited to the first order diagram $\Phi_{1}$:%
\begin{equation}%
\raisebox{-0cm}{\includegraphics[
height=0.5855in,
width=0.9703in
]%
{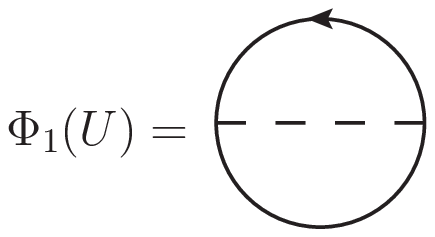}%
}%
\label{phifock}%
\end{equation}
\ We consider an expansion in powers of $e^{2}$, considering $G_{0}$, defined
in (\ref{gu0}), to be of zero order. This makes $\Phi_{1}$ proportional at
least to $e^{2}$.\ Let $e^{2}U_{1}$ be the first order approximation to
$U_{int}$:%
\begin{equation}
U_{int}\left(  x\right)  =e^{2}U_{1}\left(  x\right)
\end{equation}
Then, to first order, the equation (\ref{ueq2}) can be approximated by:%
\begin{equation}
tr\left\langle x\right|  G_{0}e^{2}U_{1}G_{0}\left|  x\right\rangle =\left.
\frac{\delta\Phi_{1}\left(  U\right)  }{\delta U\left(  x\right)  }\right|
_{U=U_{0}}\label{u1eq}%
\end{equation}
We can use (\ref{drr}) to reduce this equation to:%
\begin{equation}
-\int d^{4}x^{\prime}\;\left\langle \vec{r}\left|  D\right|  \vec{r}^{\prime
}\right\rangle e^{2}U_{1}\left(  \vec{r}^{\prime}\right)  =\frac{\delta
\Phi_{1}\left(  U_{0}\right)  }{\delta U_{0}\left(  x\right)  }\label{first1}%
\end{equation}
so that:%
\begin{equation}
e^{2}U_{1}\left(  \vec{r}\right)  =-\int d^{3}\vec{r}^{\prime}\left\langle
\vec{r}\left|  D^{-1}\right|  \vec{r}^{\prime}\right\rangle \;\frac{\delta
\Phi_{1}\left(  U_{0}\right)  }{\delta U_{0}\left(  x\right)  }\label{eu1r}%
\end{equation}
The equation (\ref{u0fromu1}) then yields the potential $U_{0}\left(  \vec
{r}\right)  $ which generates the Kohn-Sham orbits:%
\[
U_{0}\left(  \vec{r}\right)  =\int d^{3}r^{\prime}\;\frac{e^{2}}{4\pi\left|
\vec{r}-\vec{r}^{\prime}\right|  }n\left(  \vec{r}^{\prime}\right)  +\int
d^{3}\vec{r}^{\prime}\left\langle \vec{r}\left|  D^{-1}\right|  \vec
{r}^{\prime}\right\rangle \;\frac{\delta\Phi_{1}
\left(  U_{0}\right)  }{\delta
U_{0}\left(  x\right)  }%
\]%
\begin{equation}%
\raisebox{-0.0519in}{\includegraphics[
height=1.5229in,
width=2.7925in
]%
{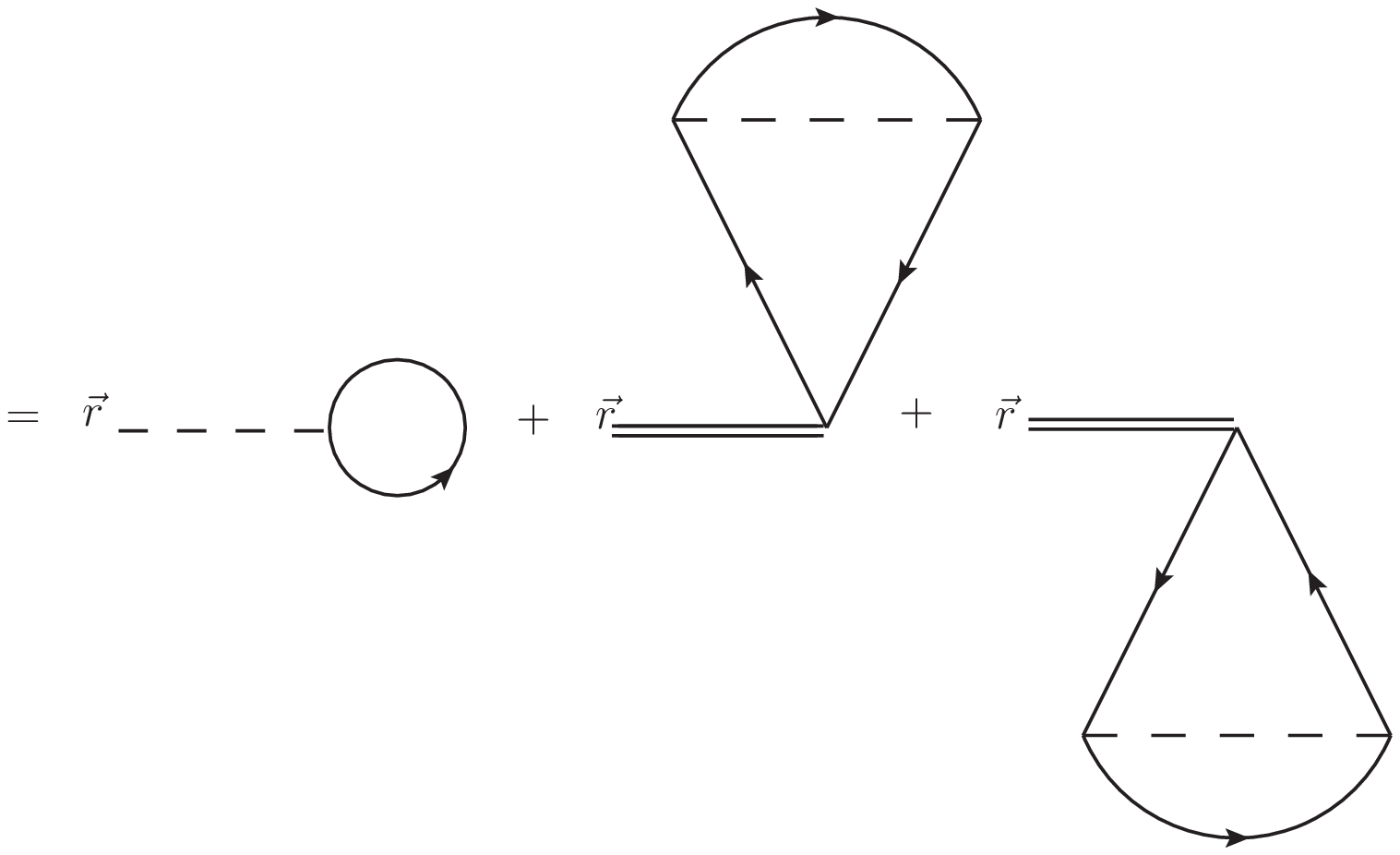}%
}%
\label{ksexch}%
\end{equation}
The diagrams in (\ref{ksexch}) are time-independent Goldstone diagrams in
which the upgoing lines are particle Kohn-Sham orbits and the downgoing line
hole orbits. The double line is the inverse density-density correlation
function (\ref{drrm1}). The first term is the Hartree potential and the second
term is the local potential which is generated by the Fock term.

To order $e^{2}$, the functional (\ref{wuint}) reduces to:%
\begin{equation}
W_{1}\left(  U\right)  =Tr\ln G_{0}^{-1}-\frac{1}{2}nKn+TrG_{0}e^{2}U_{1}%
+\Phi_{1}\left(  U_{0}\right)
\end{equation}
We can use (\ref{u12rho}) to get:%
\begin{equation}
W_{1}\left(  U_{0}\right)  =Tr\ln G_{0}^{-1}-\frac{1}{2}nKn+TrG_{0}\left(
-Kn-U_{0}\right)  +\Phi_{1}\left(  U_{0}\right)
\end{equation}
However:%
\begin{equation}
-TrG_{0}Kn=nKn\;\;\;\;\;\;\;-TrG_{0}U_{0}=nU_{0}%
\end{equation}
so that:%
\[
W_{1}\left(  U_{0}\right)  =Tr\ln G_{0}^{-1}+nU_{0}+\frac{1}{2}nKn+\Phi
_{1}\left(  U_{0}\right)
\]%
\begin{equation}%
\raisebox{-0cm}{\includegraphics[
height=0.4419in,
width=2.4379in
]%
{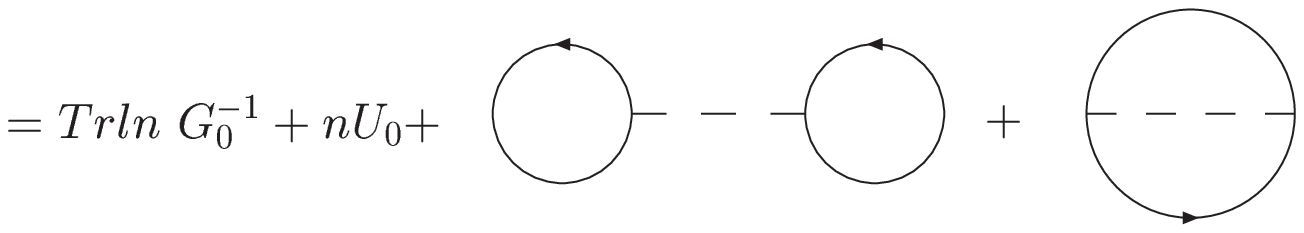}%
}%
\label{w1hf}%
\end{equation}

The first order potential (\ref{ksexch}) and functional (\ref{w1hf}) are
identical to the first order potential $J_{0}\left(  x\right)  $ 
and to the 
functional $\Gamma_{0}+\Gamma_{1}$ obtained using a Legendre transform as
given, for example, by Valiev and Fernando\cite{ValievFernando1998}.

\subsection{Second order approximation.}

Let consider the second order approximation. We limit the one-line irreducible
diagrams to the first and second order diagrams:%
\begin{equation}%
\raisebox{-0cm}{\includegraphics[
height=1.3474in,
width=2.9014in
]%
{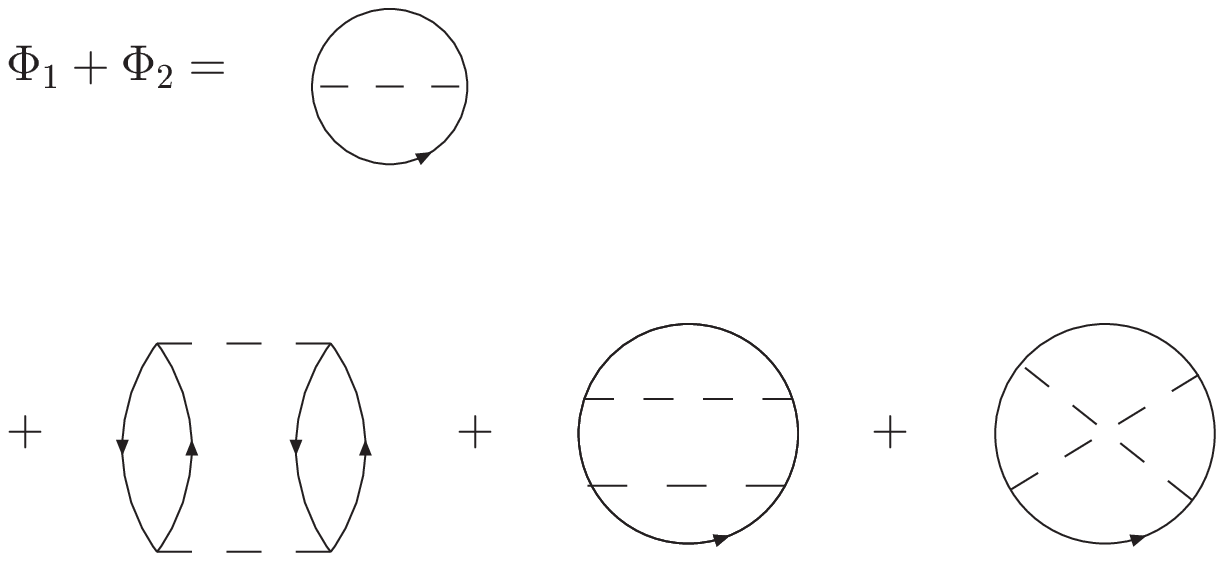}%
}%
\label{phi1phi2}%
\end{equation}
The first order diagram $\Phi_{1}$ is at least of order $e^{2}$ and the second
order diagrams at least of order $e^{4}$. The second order approximation to
$U_{int}$ can be written in the form:%
\begin{equation}
U_{int}=e^{2}U_{1}+e^{4}U_{2}%
\end{equation}
By substituting $G=G_{0}-G_{0}U_{int}G$ in the left hand side of equation
(\ref{ueq2}) it becomes:%
\begin{equation}
tr\;\left\langle x\left|  G_{0}U_{int}G_{0}\right|  x\right\rangle
=tr\;\left\langle x\left|  G_{0}U_{int}G_{0}U_{int}G\right|  x\right\rangle
+\left.  \frac{\delta\Phi\left(  U\right)  }{\delta U\left(  x\right)
}\right|  _{U=U_{0}+U_{int}} \label{ueq3}%
\end{equation}

To order $e^{2}$, the equation (\ref{ueq3}) reads:%
\begin{equation}
tr\left\langle x\right|  G_{0}e^{2}U_{1}G_{0}\left|  x\right\rangle
=\frac{\delta\Phi_{1}\left(  U_{0}\right)  }{\delta U_{0}\left(  x\right)
}\label{e2u1}%
\end{equation}
which is the result (\ref{u1eq}) obtained in first order theory, as expected.

Let us now expand the functional $W\left(  U\right)  $ up to second order.
From (\ref{wuint}), and to order $e^{4}$, we obtain:%
\[
W_{2}\left(  U\right)  =Tr\ln G_{0}^{-1}-\frac{1}{2}nKn+TrG_{0}\left(
e^{2}U_{1}+e^{4}U_{2}\right)  -\frac{1}{2}TrG_{0}\left(  e^{2}U_{1}\right)
G_{0}\left(  e^{2}U_{1}\right)
\]%
\begin{equation}
+\Phi_{1}\left(  U_{0}+e^{2}U_{1}\right)  +\Phi_{2}\left(  U_{0}\right)
\label{w2ex}%
\end{equation}
For the third term on the right hand side, use (\ref{u12rho}) to write:%
\begin{equation}
e^{2}U_{1}+e^{4}U_{2}=-\int d^{4}x^{\prime}\;\left\langle x\left|  K\right|
x^{\prime}\right\rangle n\left(  x^{\prime}\right)
\end{equation}%
\begin{equation}
TrG_{0}\left(  e^{2}U_{1}+e^{4}U_{2}\right)  =TrG_{0}\left(  -Kn\right)  =nKn
\end{equation}
so that:%
\[
W_{2}\left(  U\right)  =Tr\ln G_{0}^{-1}+\frac{1}{2}nKn-\frac{1}{2}%
TrG_{0}e^{2}U_{1}G_{0}e^{2}U_{1}%
\]%
\begin{equation}
+\Phi_{1}\left(  U_{0}+e^{2}U_{1}\right)  +\Phi_{2}\left(  U_{0}\right)
\end{equation}
We see that $W_{2}\left(  U\right)  $ depends only on $U_{1}$. The third term
of (\ref{w2ex}) is:%
\begin{equation}
-\frac{1}{2}TrG_{0}e^{2}U_{1}G_{0}e^{2}U_{1}=-\frac{1}{2}tr\int d^{4}%
x\;\left\langle x\left|  G_{0}e^{2}U_{1}G_{0}\right|  x\right\rangle
e^{2}U_{1}\left(  x\right)
\end{equation}
We can use (\ref{e2u1}) to obtain:%
\begin{equation}
-\frac{1}{2}TrG_{0}e^{2}U_{1}G_{0}e^{2}U_{1}=-\frac{1}{2}\int d^{4}%
x\;\frac{\delta\Phi_{1}\left(  U_{0}\right)  }{\delta U_{0}\left(  x\right)
}e^{2}U_{1}\left(  x\right)
\end{equation}
The before last term is, to order $e^{4}$:%
\begin{equation}
\Phi_{1}\left(  U_{0}+e^{2}U_{1}\right)  =\Phi_{1}\left(  U_{0}\right)  +\int
d^{4}x\;\frac{\delta\Phi_{1}}{\delta U_{1}\left(  x\right)  }e^{2}U_{1}\left(
x\right)
\end{equation}
Finally $W_{2}$ becomes:%
\begin{equation}
W_{2}=Tr\ln G_{0}^{-1}+\frac{1}{2}nKn+\Phi_{1}\left(  U_{0}\right)  +\Phi
_{2}\left(  U_{0}\right)  +\frac{1}{2}\int d^{4}x\;\frac{\delta\Phi_{1}%
}{\delta U_{1}\left(  x\right)  }e^{2}U_{1}\left(  x\right)  \label{w2}%
\end{equation}
Since $U_{1}$ is given by (\ref{eu1r}), the last term is:%
\begin{equation}%
\raisebox{-0cm}{\includegraphics[
height=0.5353in,
width=2.4232in
]%
{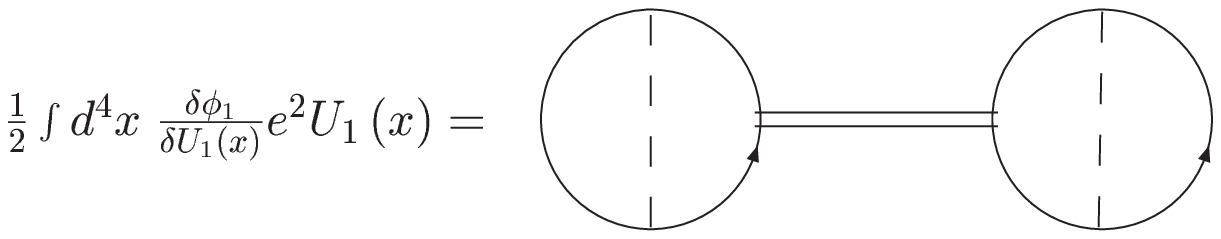}%
}%
\label{extraterm}%
\end{equation}
The functional $W_{2}$ is exactly the same as the second order functional
obtained using a Legendre transform as given, for example, by Valiev and
Fernando\cite{ValievFernando1998}.

\section{Comparison to the use of optimal orbits and conclusion.}

The potential (\ref{ksexch}) and the density functional (\ref{w2}) are
expressed in terms of Kohn-Sham orbits (\ref{ksorbs}) which are eigenstates of
$h_{0}+U_{0}$. Let us compare them to the potential and density functional
which are obtained with the optimal orbits (\ref{fermorbs}) which are
eigenstates of $h_{0}+U$. For a given set of one-line irreducible diagrams,
the potential $U\left(  x\right)  $ is given by (\ref{upotu}) as illustrated
in (\ref{uphiu}).\ When the set of irreducible diagrams is limited to the
first order diagram (\ref{phifock}), the potential $U$ is given by:%
\begin{equation}%
\raisebox{-0cm}{\includegraphics[
height=1.5229in,
width=3.3079in
]%
{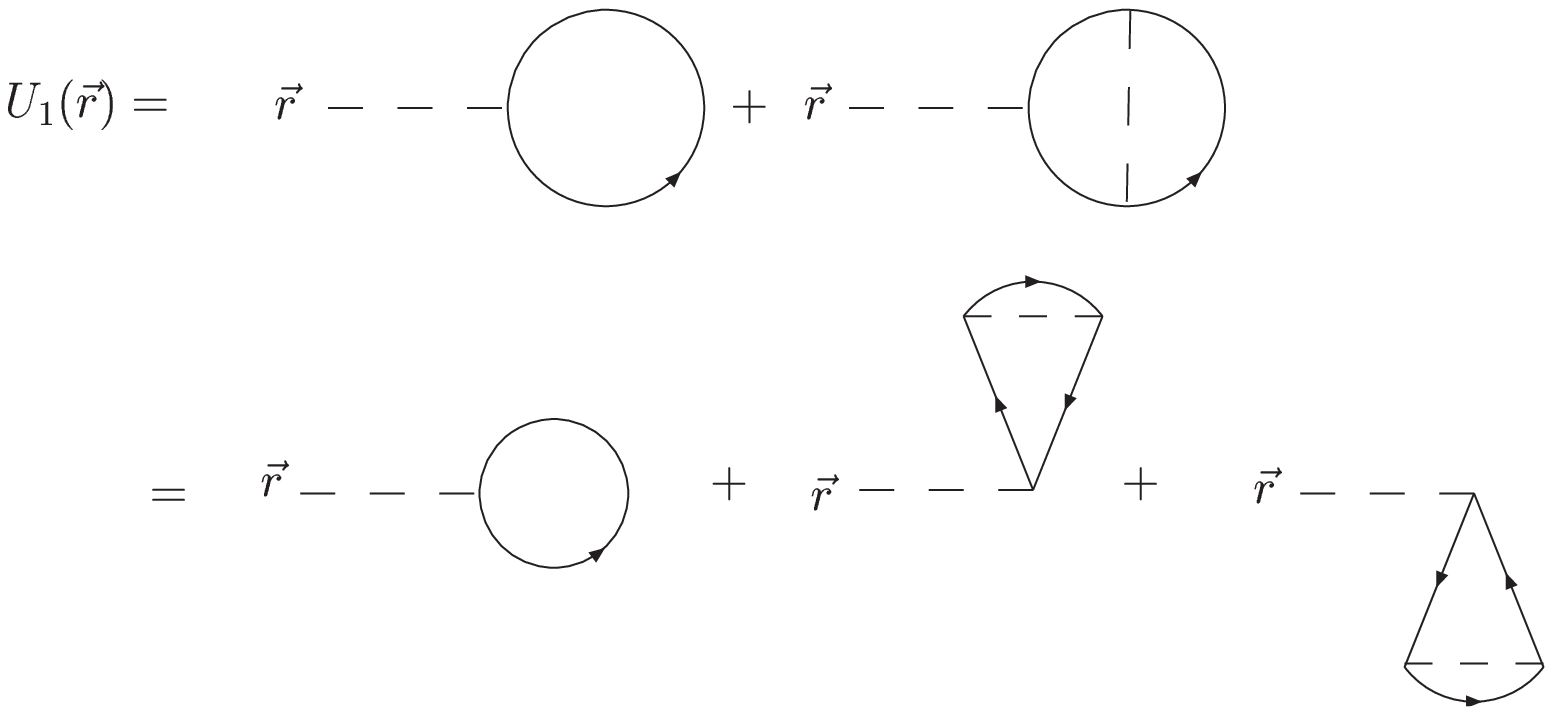}%
}%
\label{u1x}%
\end{equation}
and density $n$ is given by the diagrams illustrated in (\ref{rhoaphiu}),
namely:%
\begin{equation}%
\raisebox{-0cm}{\includegraphics[
height=1.8749in,
width=2.5668in
]%
{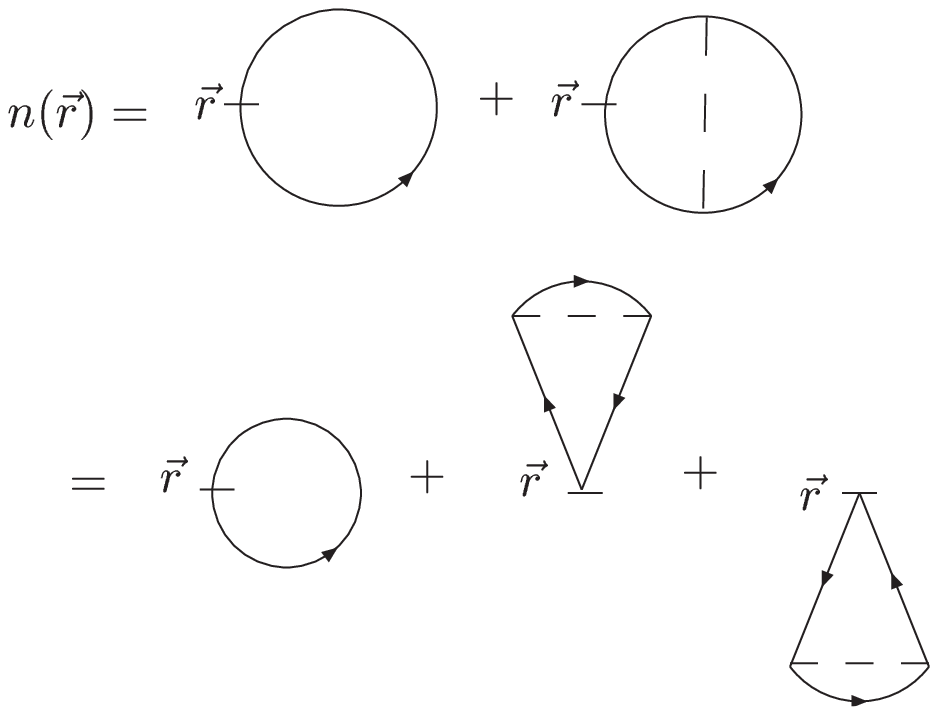}%
}%
\label{rho1x}%
\end{equation}
The first lines of the diagrams (\ref{u1x}) are Feynman diagrams whereas the
second lines are the corresponding Goldstone diagrams.

The first order potential (\ref{u1x}), expressed in terms of optimal orbits,
is simpler than the first order potential (\ref{ksexch}) expressed in terms of
Kohn-Sham orbits, because it does not involve the inverse density-density
correlation function (\ref{drr}) represented by the double line. 
But the density
(\ref{densham}) expressed in terms of Kohn-Sham orbits is simpler than the
density (\ref{rho1x}) expressed in terms of the optimal orbits. However, when
the potential is expressed in terms of optimal orbits, the density is related
to the potential by the analytic expression (\ref{upotu}), which, in the present case
of fermions with Coulomb interactions, is simply:%
\begin{equation}
n\left(  \vec{r}\right)  =-\frac{1}{e^{2}}\nabla^{2}U\left(  \vec{r}\right)
\end{equation}
Therefore no extra work is required to calculate the density $n\left(  \vec
{r}\right)  $ once the potential $U\left(  \vec{r}\right)  $ has been
calculated. Furthermore, the second order functional (\ref{w2}), expressed in
terms of Kohn-Sham orbits, has the extra contribution (\ref{extraterm}) which
does not occur in the second order functional expressed in terms of optimal
orbits. When optimal orbits are used, the simple relation (\ref{upotu})
between the potential $U$ and the density $n$ avoids having to invert the
equation $n\left(  x\right)  =-\frac{\delta W}{\delta J\left(  x\right)  }$ in
the Legendre transform method. Furthermore, any (finite or infinite) subset
of one-line irreducible diagrams can be chosen. The use of Kohn-Sham orbits
appears to be an unnecessary complication.

\appendix

\renewcommand{\theequation}{\Alph{section}.\arabic{equation}}

\setcounter{equation}{0}

\section{Diagram rules for fermions with Coulomb interactions.}

\label{ap:diagclb}

We summarize the Feynman rules in order to fix the diagram notation.\ A more
detailed derivation can be found in textbooks \cite{Ripka86},\cite{Negele1988}.

\subsection{Expansion of the partition function in terms of Feynman diagrams.}

We add a source term $j$ for the boson field $\omega_{0}$ so that the
euclidean action (\ref{actclb}) becomes:%
\[
I_{j}\left(  \omega_{0}\right)  =-Tr\ln\left(  \partial_{\tau}+h_{0}%
-\mu+i\omega_{0}\right)
\]%
\begin{equation}
-\frac{1}{2}\int d^{4}x_{1}d^{4}x_{2}\;\omega_{0}\left(  x_{1}\right)
\left\langle x_{1}\left|  K^{-1}\right|  x_{2}\right\rangle \omega_{0}\left(
x_{2}\right)  -\int d^{4}x\;j\left(  x\right)  \omega_{0}\left(  x\right)
\end{equation}
and we write the partition function (\ref{partclb}) in the form:%
\[
e^{W\left(  j\right)  }=\int D\left(  \omega_{0}\right)  e^{-I_{j}\left(
\omega_{0}\right)  }%
\]
We define an unperturbed partition function in obvious short hand notation:%
\begin{equation}
e^{W_{0}\left(  j\right)  }=e^{Tr\ln\left(  \partial_{\tau}+h_{0}-\mu\right)
}\int D\left(  \omega_{0}\right)  \;e^{\frac{1}{2}\omega_{0}K^{-1}\omega
_{0}+j\omega_{0}}%
\end{equation}
so that:%
\begin{equation}
W_{0}\left(  j\right)  =Tr\ln\left(  \partial_{\tau}+h_{0}-\mu\right)
-\frac{1}{2}jKj=Tr\ln g^{-1}-\frac{1}{2}jKj\label{w0clb}%
\end{equation}
where $g$ is the unperturbed fermion propagator:%
\begin{equation}
g=\frac{1}{\partial_{\tau}+h_{0}-\mu}\label{smallgclb}%
\end{equation}
It is the quantity:%
\begin{equation}
e^{-W_{0}\left(  j\right)  }e^{W\left(  j\right)  }=\frac{\int D\left(
\omega_{0}\right)  e^{\frac{1}{2}\omega_{0}K^{-1}\omega_{0}+j\omega_{0}%
}e^{Tr\ln\left(  1+gi\omega_{0}\right)  }}{\int D\left(  \omega_{0}\right)
e^{\frac{1}{2}\omega_{0}K^{-1}\omega_{0}+j\omega_{0}}}\equiv\left\langle
e^{Tr\ln\left(  1+gi\omega_{0}\right)  }\right\rangle \label{ewzclb}%
\end{equation}
which we expand in terms of Feynman diagrams. In the expression (\ref{ewzclb})
$\left\langle {}\right\rangle $ signifies an integration over the boson field
$\omega_{0}$ according to the expression above.

The expansion of $Tr\ln\left(  1+gi\omega_{0}\right)  $ reads:%
\[
Tr\ln\left(  1+gi\omega_{0}\right)  =Tr\;gi\omega_{0}-\frac{1}{2}%
Tr\;gi\omega_{0}gi\omega_{0}+...
\]%
\begin{equation}
=\sum_{n=1}^{\infty}\left(  -\right)  ^{n+1}\frac{1}{n}\int d^{4}x_{1}%
...d^{4}x_{n}\;tr\;\left\langle x_{1}\left|  g\right|  x_{2}\right\rangle
i\omega_{0}\left(  x_{2}\right)  \left\langle x_{2}\left|  g\right|
x_{3}\right\rangle i\omega_{0}\left(  x_{3}\right)  ...\left\langle
x_{n}\left|  g\right|  x_{1}\right\rangle i\omega_{0}\left(  x_{1}\right)
\label{lpclb}%
\end{equation}
The traces $Tr$ and $tr$ are defined in (\ref{trace1}) and (\ref{trace2}). 
The term of order $n$ gives rise to a closed loop formed by $n$ fermion
propagators $\left\langle x\left|  g\right|  x^{\prime}\right\rangle $
represented by oriented lines the end points of which are vertices:%
\begin{equation}%
{\includegraphics[
height=2.3021in,
width=2.9014in
]%
{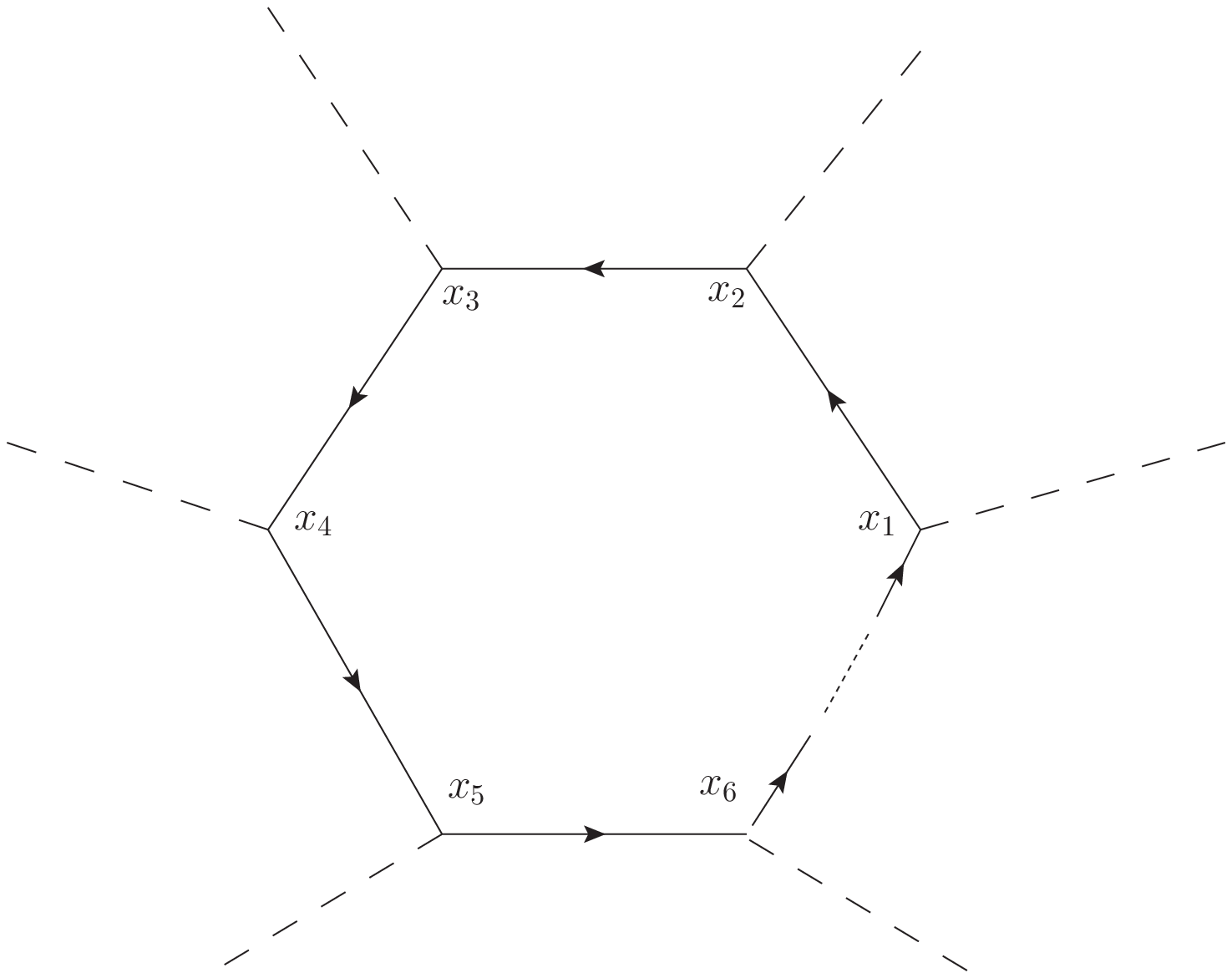}%
}%
\label{loopclb}%
\end{equation}
An oriented line which reaches a point labelled $x$ and which stems from a
point labelled $x^{\prime}$ contributes to a diagram the factor:%
\begin{equation}%
{\includegraphics[
height=0.2638in,
width=1.9173in
]%
{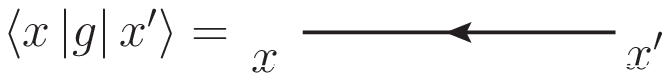}%
}%
\label{gclb}%
\end{equation}
When the expansion (\ref{lpclb}) is inserted into the expression
(\ref{ewzclb}) and the exponential $e^{Tr\ln\left(  1+gi\omega_{0}\right)  }$
is expanded in turn, one is left with expectation values of products of
boson fields:%
\[
\left\langle i\omega_{0}\left(  x_{1}\right)  i\omega_{0}\left(  x_{2}\right)
i\omega_{0}\left(  x_{3}\right)  ...\right\rangle =\frac{\int D\left(
\omega_{0}\right)  e^{\frac{1}{2}\omega_{0}K^{-1}\omega_{0}+j\omega_{0}%
}\left(  i\omega_{0}\left(  x_{1}\right)  i\omega_{0}\left(  x_{2}\right)
i\omega_{0}\left(  x_{3}\right)  ...\right)  }{\int D\left(  \omega
_{0}\right)  e^{\frac{1}{2}\omega_{0}K^{-1}\omega_{0}+j\omega_{0}}}%
\]
The expectation value of a single field is:%
\[
\left\langle i\omega_{0}\left(  x\right)  \right\rangle =i\frac{\delta
W_{0}\left(  j\right)  }{\delta j\left(  x\right)  }=-i\int d^{4}x^{\prime
}\left\langle x\left|  K\right|  x^{\prime}\right\rangle j\left(  x^{\prime
}\right)
\]%
\begin{equation}%
{\includegraphics[
height=0.3572in,
width=1.67in
]%
{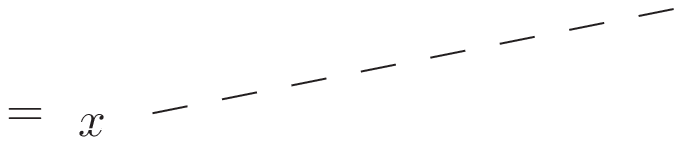}%
}%
\label{clbdash}%
\end{equation}
The expectation value (contraction) of a product of two fields is:%
\begin{equation}
\left\langle i\omega_{0}\left(  x\right)  i\omega_{0}\left(  x^{\prime
}\right)  \right\rangle =\left\langle i\omega_{0}\left(  x\right)
\right\rangle \left\langle i\omega_{0}\left(  x^{\prime}\right)  \right\rangle
+\left\langle i\omega_{0}\left(  x\right)  i\omega_{0}\left(  x^{\prime
}\right)  \right\rangle _{C}%
\end{equation}
and the connected part is the boson propagator:%
\[
\left\langle i\omega_{0}\left(  x\right)  i\omega_{0}\left(  x^{\prime
}\right)  \right\rangle _{C}=-\frac{\delta^{2}W_{0}\left(  j\right)  }{\delta
j\left(  x\right)  \delta j\left(  x^{\prime}\right)  }%
\]%
\begin{equation}%
{\includegraphics[
height=0.2733in,
width=2.8798in
]%
{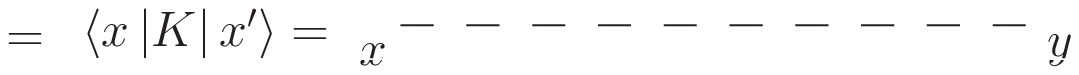}%
}%
\label{clbprop}%
\end{equation}
Note that the dashed interaction line yields a factor $K$ which is equal to
\emph{minus }the instantaneous Coulomb potential, defined in (\ref{coulpot}%
). Expectation values of products of three or more fields can be expressed in
terms (\ref{clbdash}) and (\ref{clbprop}), using Wick's theorem.

Then $e^{-W_{0}\left(  j\right)  }e^{W\left(  j\right)  }$ is equal to the sum
of all distinct unlabeled diagrams formed by closed loops of oriented fermion
lines and of dashed interaction lines.\ The contribution of a diagram is
obtained by the following rules:

\begin{itemize}
\item  Assign a label $x=\left(  \tau,\vec{r}\right)  $ to each vertex.

\item  Each oriented fermion line contributes a fermion propagator
(\ref{gclb}). The fermion propagators are written in the order in which they
appear as one follows the closed loop. For each closed loop we take a trace
$tr$ of the product of the fermion propagators forming the closed loop.

\item  Each dashed interaction line joining two vertices contributes one of
the factors (\ref{clbdash}) or (\ref{clbprop}), depending on whether it joins
one or two vertices.

\item  A diagram containing $n_{v}$ vertices and $n_{l}$ closed loops is
multiplied by the factor $\left(  -\right)  ^{n_{v}+n_{l}}$. (Note that in the
absence of the source $j$, the number of vertices is always even.)

\item  Integrate over the labels $x=\left(  \tau,\vec{r}\right)  $ of the
vertices and divide by the symmetry factor, which is equal to the number of
permutations of the labels which lead to an identical unlabeled diagram.
\end{itemize}

The contributions of disconnected diagrams factor and their symmetry factors
are such that the sum of all diagrams is equal to the exponential of the sum
$\Gamma_{c}$ of connected diagrams. It follows that the partition function can
be expressed in terms of connected diagrams $\Gamma_{c}$ as follows:%
\begin{equation}
W\left(  j\right)  =W_{0}\left(  j\right)  +\Gamma_{c}\label{conclbj}%
\end{equation}
where $W_{0}\left(  j\right)  $ is given by (\ref{w0clb}). In the absence of
the sources $j$, the partition function is given by:%
\begin{equation}
\ln\left(  Tre^{-\beta\left(  H-\mu N\right)  }\right)  =W=Tr\ln g^{-1}%
+\Gamma_{c}\label{conclb}%
\end{equation}
where $g$ is the fermion propagator (\ref{smallgclb}). An approximation to the
partition can be defined by a choice of a subset of connected diagrams
$\Gamma_{c}$.

\subsection{Diagram expansion of the particle density.}

\label{ap:denclb}

Let us add a local source term $U\left(  x\right)  $ to $h_{0}$. From
(\ref{partclbpsi} and (\ref{actnn}) we see that, upon a variation $U\left(
x\right)  \rightarrow U\left(  x\right)  +\delta U\left(  x\right)  $, we have
$\delta\ln Z=\delta W=-\int d^{4}x\;\left\langle \psi^{\dagger}\left(
x\right)  \delta U\left(  x\right)  \psi\left(  x\right)  \right\rangle $ so
that the particle density is given by:%
\begin{equation}
n\left(  x\right)  \equiv\left\langle \psi^{\dagger}\left(  x\right)
\psi\left(  x\right)  \right\rangle =-\left.  \frac{\delta W}{\delta U\left(
x\right)  }\right|  _{U=0}\label{denclb}%
\end{equation}
When $h_{0}$ is time independent, as assumed in this work, the particle
density is also time independent:%
\begin{equation}
\left\langle \psi^{\dagger}\left(  x\right)  \psi\left(  x\right)
\right\rangle =\left\langle \psi^{\dagger}\left(  \vec{r}\right)  \psi\left(
\vec{r}\right)  \right\rangle \;\;\;\;\;\;\;n\left(  x\right)  =n\left(
\vec{r}\right)
\end{equation}
However, for any time independent functional $F$, we have $\int d^{4}x\;\delta
U\left(  x\right)  \frac{\delta F}{\delta U\left(  x\right)  }=\beta\int
d^{3}r\;\delta U\left(  \vec{r}\right)  \frac{\delta F}{\delta U\left(
\vec{r}\right)  }$. In order to avoid cumbersome $\beta$ factors, we shall
always use functional derivatives $\frac{\delta}{\delta U\left(  x\right)  }$.

In the presence of the source, the fermion propagator (\ref{smallgclb})
becomes:%
\begin{equation}
g=\frac{1}{\partial_{\tau}+h_{0}-\mu+U}\label{guclb}%
\end{equation}
From the expression (\ref{conclb}) of the partition function, we see that the
diagram expansion of the particle density (\ref{denclb}) is given by:%
\begin{equation}
n\left(  x\right)  \equiv\left\langle \psi^{\dagger}\left(  x\right)
\psi\left(  x\right)  \right\rangle =-\left.  \frac{\delta}{\delta U\left(
x\right)  }\left(  Tr\ln g^{-1}+\Gamma_{c}\right)  \right|  _{U=0}%
\label{denclb1}%
\end{equation}
In (\ref{guclb}), $U$ is the following operator acting in the Hilbert space of
a single fermion:%
\begin{equation}
U=\int d^{4}x\;\left|  x\right\rangle U\left(  x\right)  \left\langle
x\right|
\end{equation}
It follows that:%
\[
\frac{\delta}{\delta U\left(  x\right)  }\left\langle x_{1}\left|  g\right|
x_{2}\right\rangle =-\left\langle x_{1}\left|  g\right|  x\right\rangle
\left\langle x\left|  g\right|  x_{2}\right\rangle
\]%
\begin{equation}
\frac{\delta}{\delta U\left(  x\right)  }Tr\;\ln g^{-1}=Tr\;g\left|
x\right\rangle \left\langle x\right|  =tr\;\left\langle x\left|  g\right|
x\right\rangle \label{dgdux}%
\end{equation}
The first term of (\ref{denclb1}) is the unperturbed particle density:%
\[
n_{0}\left(  x\right)  =-tr\;\left\langle x\left|  g\right|  x\right\rangle
=\sum_{h}\left\langle \left.  h\right|  \vec{r}\right\rangle \left\langle
\left.  \vec{r}\right|  h\right\rangle
\]%
\begin{equation}%
{\includegraphics[
height=0.518in,
width=0.8112in
]%
{nzeror.eps}%
}%
\label{nzeror}%
\end{equation}
where we used (\ref{g12zerot}). For the second term of (\ref{denclb1}) we note
that the potential $U$ occurs only in the propagators $g$ of the 
diagrams $\Gamma_{c}$ so that:%
\begin{equation}
\frac{\delta\Gamma_{c}}{\delta U\left(  x\right)  }=\int d^{4}x_{1}d^{4}%
x_{2}\;\frac{\delta\Gamma_{c}}{\delta\left\langle x_{1}\left|  g\right|
x_{2}\right\rangle }\frac{\delta\left\langle x_{1}\left|  g\right|
x_{2}\right\rangle }{\delta U\left(  x\right)  }=-\int d^{4}x_{1}d^{4}%
x_{2}\;\frac{\delta\Gamma_{c}}{\delta\left\langle x_{1}\left|  g\right|
x_{2}\right\rangle }\left\langle x_{1}\left|  g\right|  x\right\rangle
\left\langle x\left|  g\right|  x_{2}\right\rangle
\end{equation}
The second term is the sum of distinct diagrams obtained from the unlabeled
connected diagrams $\Gamma_{c}$ by inserting a slash, labelled $x$, onto its
oriented fermion lines. The particle density can therefore be expressed in
terms of the connected diagrams thus:%
\begin{equation}
n\left(  x\right)  =-tr\;\left\langle x\left|  g\right|  x\right\rangle +\int
d^{4}x_{1}d^{4}x_{2}\;\frac{\delta\Gamma_{c}}{\delta\left\langle x_{1}\left|
g\right|  x_{2}\right\rangle }\left\langle x_{1}\left|  g\right|
x\right\rangle \left\langle x\left|  g\right|  x_{2}\right\rangle
\label{partden}%
\end{equation}
\ For example, if the only retained connected diagrams are:%
\begin{equation}%
{\includegraphics[
height=0.7247in,
width=3.6798in
]%
{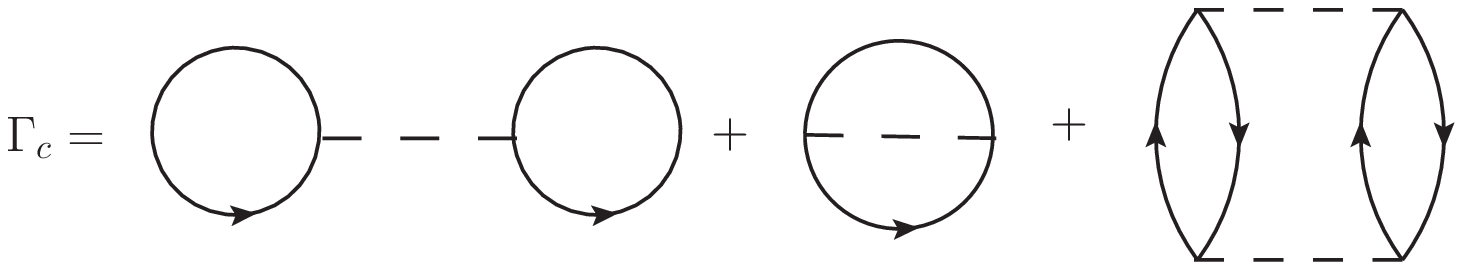}%
}%
\label{retclb}%
\end{equation}
the particle density becomes:%
\begin{equation}%
{\includegraphics[
height=0.9729in,
width=4.8896in
]%
{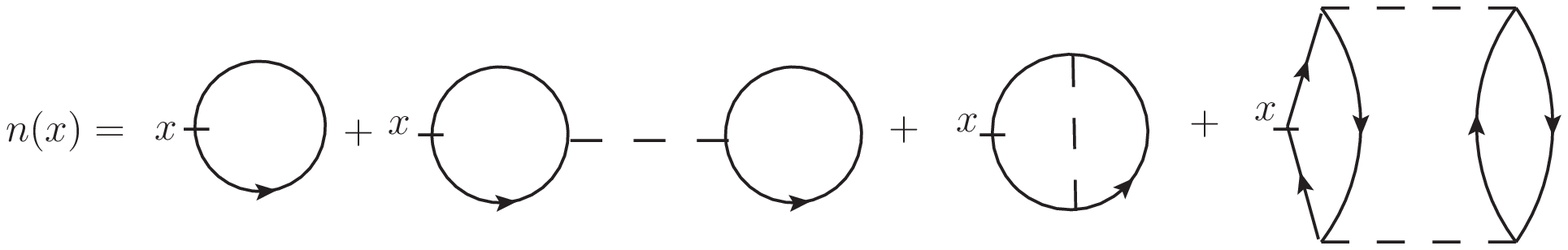}%
}%
\label{retdiag}%
\end{equation}
The diagrams (\ref{retdiag}) have a slash, labelled $x$ which does not modify
the sign of the diagram.

\renewcommand{\theequation}{\Alph{section}.\arabic{equation}}

\setcounter{equation}{0}

\section{Diagram rules for nucleons interacting with mesons.}

\label{ap:nucmesdiag}

\subsection{Expansion of the partition function in terms of Feynman diagrams.}

For a system of nucleons interacting with several mesons, 
we add to $h_0$ the following
source terms for the boson fields:
\begin{equation}
jS=\int d^{4}x\left[  j_{\sigma}\left(  x\right)  \sigma\left(  x\right)
+j_{a}\left(  x\right)  \pi_{a}\left(  x\right)  +j_{\mu a}\left(  x\right)
\left(  \partial_{\mu}\pi_{a}\right)  _{x}+j_{\mu}\left(  x\right)
\omega_{\mu}\left(  x\right)  \right]
\end{equation}
The euclidean action (\ref{actphi}) becomes:%
\begin{equation}
I_{j}\left(  S\right)  \equiv-Tr\ln\beta\left(  \partial_{\tau}+h_{0}%
-\mu+\left(  \Gamma S\right)  +S\left(  \Theta S\right)  \right)  +\frac{1}%
{2}SK^{-1}S-jS
\end{equation}
and we write the partition function (\ref{partfun}) in the form:%
\begin{equation}
e^{W\left(  j\right)  }=\int D\left(  S\right)  e^{-I_{j}\left(  S\right)  }%
\end{equation}
The unperturbed partition function is obtained by setting to zero the coupling
constants so that $\left(  \Gamma S\right)  =S\left(  \Theta S\right)
=0$.\ In obvious short hand notation:%
\begin{equation}
e^{W_{0}\left(  j\right)  }=e^{Tr\ln\left(  \partial_{\tau}+h_{0}-\mu\right)
}\int D\left(  S\right)  e^{-\frac{1}{2}SK^{-1}S+jS}=e^{Tr\ln\left(
\partial_{\tau}+h_{0}-\mu\right)  +\frac{1}{2}jKj}%
\end{equation}
so that:%
\begin{equation}
W_{0}\left(  j\right)  =Tr\ln\left(  \partial_{\tau}+h_{0}-\mu\right)
+\frac{1}{2}jKj=Tr\ln g^{-1}+\frac{1}{2}jKj\label{wzero2}%
\end{equation}
where $g$ is the unperturbed fermion propagator:
\begin{equation}
g=\frac{1}{\partial_{\tau}+h_{0}-\mu}\label{gprop}%
\end{equation}
In more explicit form, the expression (\ref{wzero2}) reads:%
\[
W_{0}\left(  j\right)  =Tr\ln g^{-1}%
\]%
\begin{equation}
+\frac{1}{2}j_{\sigma}\frac{1}{-\partial^{2}+m_{\sigma}^{2}}j_{\sigma}%
+\frac{1}{2}j_{\pi a}\frac{1}{-\partial^{2}+m_{\pi}^{2}}j_{\pi a}+\frac{1}%
{2}j_{\omega\mu}\frac{1}{\left(  -\partial^{2}+m_{\omega}^{2}\right)  }\left(
\delta_{\mu\nu}-\frac{1}{m^{2}}\partial_{\mu}\partial_{\nu}\right)
j_{\omega\nu}\label{wzeroj}%
\end{equation}
The traces $Tr$ and $tr$ are defined in equation (\ref{trace}).

It is the quantity:%
\begin{equation}
e^{-W_{0}\left(  j\right)  }e^{W\left(  j\right)  }=\frac{\int D\left(
S\right)  e^{Tr\ln\left(  1+g\left(  \Gamma S\right)  +gS\left(  \Phi
S\right)  \right)  }e^{-\frac{1}{2}SK^{-1}S+jS}}{\int D\left(  S\right)
e^{-\frac{1}{2}SK^{-1}S+jS}}\equiv\left\langle e^{Tr\ln\left(  1+g\left(
\Gamma S\right)  +gS\left(  \Phi S\right)  \right)  }\right\rangle
\label{ewzew}%
\end{equation}
which we express in terms of Feynman diagrams. In the expression
(\ref{ewzew}), $\left\langle {}\right\rangle $ signifies an integration over
the boson fields $S$ according to the expression above.

The expansion of $Tr\ln\left(  1+g\left(  \Gamma S\right)  +gS\left(  \Phi
S\right)  \right)  $ in powers of $\left(  \Gamma S\right)  +S\left(  \Phi
S\right)  $ reads:%
\[
Tr\ln\left(  1+g\left(  \Gamma S\right)  +gS\left(  \Phi S\right)  \right)
\]%
\[
=Trg\left[  \left(  \Gamma S\right)  +S\left(  \Phi S\right)  \right]
-\frac{1}{2}Trg\left[  \left(  \Gamma S\right)  +S\left(  \Phi S\right)
\right]  g\left[  \left(  \Gamma S\right)  +S\left(  \Phi S\right)  \right]
+...
\]%
\[
=\sum_{n=1}^{\infty}\left(  -1\right)  ^{n+1}\frac{1}{n}\int d^{4}x_{1}%
d^{4}x_{2}...d^{4}x_{n}\;tr\left\{  \left\langle x_{1}\left|  g\right|
x_{2}\right\rangle \left[  \left(  \Gamma S_{x_{2}}\right)  +S_{x_{2}}\left(
\Phi S_{x_{2}}\right)  \right]  \right.
\]%
\begin{equation}
\left.  \left\langle x_{2}\left|  g\right|  x_{3}\right\rangle \left[  \left(
\Gamma S_{x_{3}}\right)  +S_{x_{3}}\left(  \Phi S_{x_{3}}\right)  \right]
...\left\langle x_{n}\left|  g\right|  x_{1}\right\rangle \left[  \left(
\Gamma S_{x_{1}}\right)  +S_{x_{1}}\left(  \Phi S_{x_{1}}\right)  \right]
\right\}  \label{floop}%
\end{equation}
The term of order $n$ is a trace $tr$ of a product of terms the explicit form
of which is:%
\[
\left\langle x_{1}\left|  g\right|  x_{2}\right\rangle \left[  \left(  \Gamma
S_{x_{2}}\right)  +S_{x_{2}}\left(  \Phi S_{x_{2}}\right)  \right]
\]%
\begin{equation}
=\left\langle x_{1}\left|  \frac{1}{\partial_{\tau}+h_{0}-\mu}\right|
x_{2}\right\rangle \left[  \Gamma^{\left(  \sigma\right)  }\sigma\left(
x_{2}\right)  +\Gamma_{\mu}^{\left(  \omega\right)  }\omega_{\mu}\left(
x_{2}\right)  +\Gamma_{\mu a}^{\left(  \pi\right)  }\left(  \partial_{\mu}%
\pi_{a}\right)  _{x_{2}}+\pi_{a}\left(  x_{2}\right)  \Phi_{a\left(  \mu
b\right)  }^{\left(  \pi\right)  }\left(  \partial_{\mu}\pi_{b}\right)
_{x_{2}}\right]
\end{equation}
The product can be represented by a diagram of the form:%
\begin{equation}%
{\includegraphics[
height=4.024in,
width=5.1647in
]%
{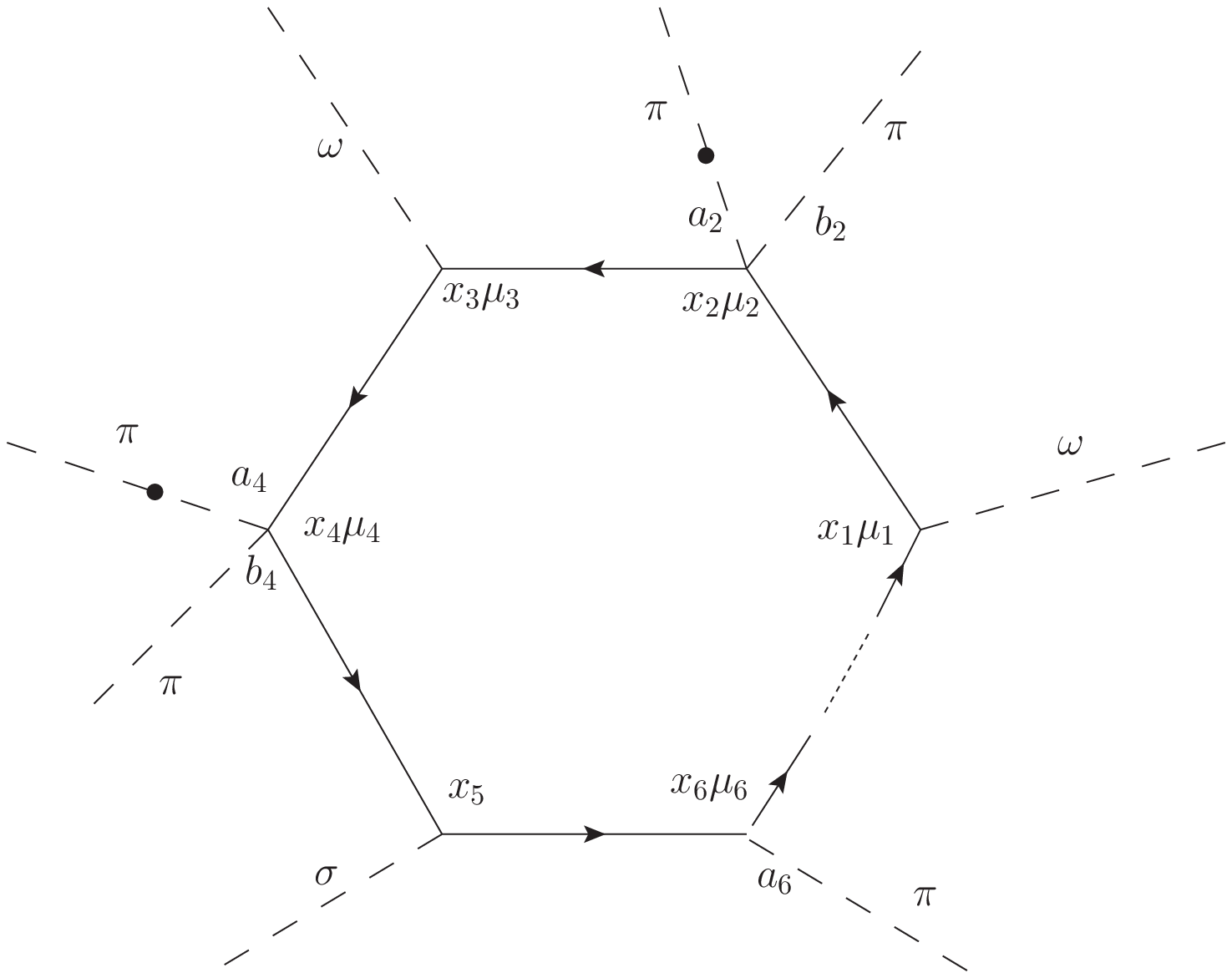}%
}%
\label{loop}%
\end{equation}
in which the oriented (fermion propagator) lines form a closed loop. Each
oriented fermion propagator contributes one of the following factors:%
\[%
{\includegraphics[
height=0.9236in,
width=1.8593in
]%
{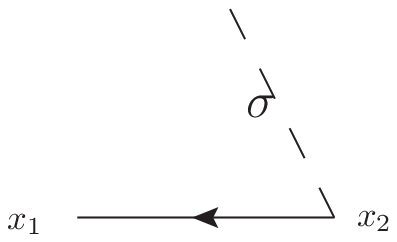}%
}%
\]%
\[
=\left\langle x_{1}\left|  g\right|  x_{2}\right\rangle \Gamma^{\left(
\sigma\right)  }=\left\langle x_{1}\left|  \frac{1}{\partial_{\tau}+h_{0}-\mu
}\right|  x_{2}\right\rangle g_{\sigma}\beta
\]%
\[%
{\includegraphics[
height=0.9582in,
width=1.9285in
]%
{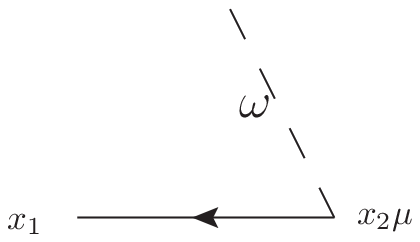}%
}%
\]%
\[
=\left\langle x_{1}\left|  g\right|  x_{2}\right\rangle \Gamma_{\mu}^{\left(
\omega\right)  }=\left\langle x_{1}\left|  \frac{1}{\partial_{\tau}+h_{0}-\mu
}\right|  x_{2}\right\rangle g_{\omega}\beta\gamma_{\mu}%
\]%
\[%
{\includegraphics[
height=1.0058in,
width=2.0245in
]%
{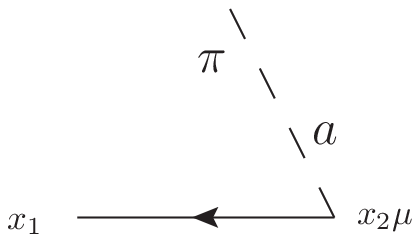}%
}%
\]%
\[
=\left\langle x_{1}\left|  g\right|  x_{2}\right\rangle \Gamma_{\mu
a}^{\left(  \pi\right)  }=\left\langle x_{1}\left|  \frac{1}{\partial_{\tau
}+h_{0}-\mu}\right|  x_{2}\right\rangle \frac{g_{A}}{2f_{\pi}}\beta\tau
_{a}\gamma_{5}\gamma_{\mu}%
\]%
\[%
{\includegraphics[
height=1.0646in,
width=2.0072in
]%
{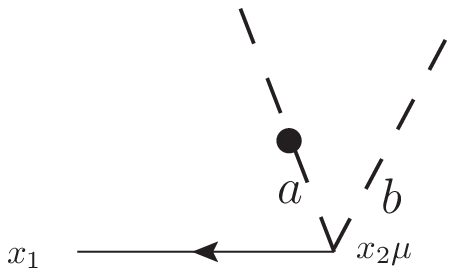}%
}%
\]%
\begin{equation}
=\left\langle x_{1}\left|  g\right|  x_{2}\right\rangle \Phi_{a(\mu
b)}=\left\langle x_{1}\left|  \frac{1}{\partial_{\tau}+h_{0}-\mu}\right|
x_{2}\right\rangle \frac{1}{4f_{\pi}^{2}}\varepsilon_{abc}\beta\tau_{c}%
\gamma_{\mu}\label{twopivert}%
\end{equation}
Note that the order of the non-commuting operators $g$ and $\left(  \Gamma
S\right)  $ or $S\left(  \Phi S\right)  $ is determined by the direction of
the arrow on the fermion propagator. The point at which a meson (dashed) line
joins a fermion (oriented) line is called a vertex. The proper labelling of
the vertices is important. Note also that \emph{the black dot denotes the
absence} of a derivative of the pion field.

When the expansion (\ref{floop}) is inserted into the expression (\ref{ewzew})
and the exponential $e^{Tr\ln\left(  1+g\left(  \Gamma S\right)  +gS\left(
\Phi S\right)  \right)  }$ is expanded in turn, we are left with 
expectation values of products of boson fields which can be evaluated using
Wick's theorem. The expectation values of single fields are:%
\begin{equation}
\left\langle S_{i}\left(  x\right)  \right\rangle =\int d^{4}y\;\left\langle
x\left|  K_{ij}\right|  y\right\rangle \;j_{j}\left(  y\right)  =\left\langle
x\left|  K_{ij}\right|  j_{j}\right\rangle \label{sax}%
\end{equation}
In more explicit form:%
\[
\left\langle \sigma\left(  x\right)  \right\rangle =\frac{\partial W_{0}%
}{\partial j_{\sigma}\left(  x\right)  }=\int d^{4}y\;\left\langle x\left|
\frac{1}{-\partial^{2}+m_{\sigma}^{2}}\right|  y\right\rangle j_{\sigma
}\left(  y\right)  =\left\langle x\left|  K^{\left(  \sigma\right)  }\right|
j_{\sigma}\right\rangle
\]%
\[%
\raisebox{-0cm}{\includegraphics[
height=0.4782in,
width=1.8014in
]%
{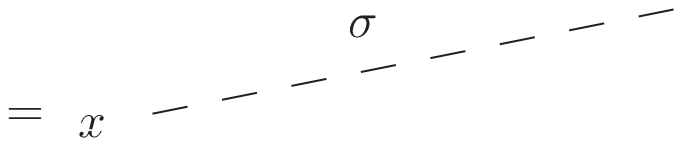}%
}%
\]%
\[
\left\langle \pi_{a}\left(  x\right)  \right\rangle =\frac{\partial W_{0}%
}{\partial j_{\pi a}}=\int d^{4}y\;\left\langle x\left|  \frac{1}%
{-\partial^{2}+m_{\pi}^{2}}\right|  y\right\rangle j_{\pi a}\left(  y\right)
=\left\langle x\left|  K^{\left(  \pi\right)  }\right|  j_{\pi a}%
\right\rangle
\]%
\[%
\raisebox{-0cm}{\includegraphics[
height=0.4255in,
width=1.7582in
]%
{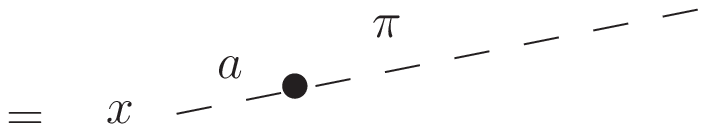}%
}%
\]%
\[
\left\langle \partial_{\mu}\pi_{a}\left(  x\right)  \right\rangle =\int
d^{4}y\;\left\langle x\left|  \partial_{\mu}\frac{1}{-\partial^{2}+m_{\pi}%
^{2}}\right|  y\right\rangle j_{\pi a}\left(  y\right)  =\left\langle x\left|
\partial_{\mu}K^{\left(  \pi\right)  }\right|  j_{\pi a}\right\rangle
\]%
\[%
\raisebox{-0cm}{\includegraphics[
height=0.4255in,
width=1.823in
]%
{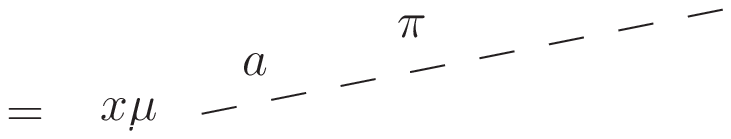}%
}%
\]%
\[
\left\langle \omega_{\mu}\left(  x\right)  \right\rangle =\frac{\partial
W_{0}}{\partial j_{\omega\mu}}%
\]%
\[
=\int d^{4}y\;\left\langle x\left|  \frac{1}{\left(  -\partial^{2}+m_{\omega
}^{2}\right)  }\left(  \delta_{\mu\nu}-\frac{1}{m_{\omega}^{2}}\partial_{\mu
}\partial_{\nu}\right)  \right|  y\right\rangle j_{\omega\nu}\left(  y\right)
=\left\langle x\left|  K_{\mu\nu}^{\left(  \omega\right)  }\right|
j_{\omega\nu}\right\rangle
\]%
\begin{equation}%
\raisebox{-0cm}{\includegraphics[
height=0.4393in,
width=1.8135in
]%
{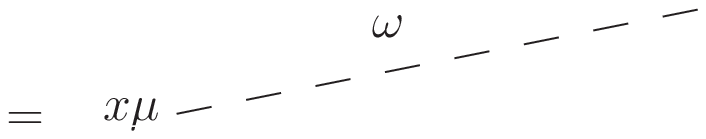}%
}%
\label{omegaxmu}%
\end{equation}
The expectation values of products of two fields are:%
\begin{equation}
\left\langle S_{a}\left(  x\right)  S_{b}\left(  y\right)  \right\rangle
=e^{-W_{0}\left(  j\right)  }\frac{\delta^{2}}{\delta j_{a}\left(  x\right)
\delta j_{b}\left(  y\right)  }e^{W_{0}\left(  j\right)  }=\left\langle
S_{i}\left(  x\right)  \right\rangle \left\langle S_{j}\left(  y\right)
\right\rangle +\left\langle S_{i}\left(  x\right)  S_{j}\left(  y\right)
\right\rangle _{c}%
\end{equation}
where the connected part is the boson propagator:%
\begin{equation}
\left\langle S_{a}\left(  x\right)  S_{b}\left(  y\right)  \right\rangle
_{c}=\frac{\delta^{2}W_{0}\left(  j\right)  }{\delta j_{a}\left(  x\right)
\delta j_{b}\left(  y\right)  }=\left\langle x\left|  K_{ab}\right|
y\right\rangle \label{sabxy}%
\end{equation}
where $K_{ab}$ is the meson propagator (\ref{kab}). We can use (\ref{wzeroj})
to deduce the following explicit forms of the boson propagators:%
\[
\left\langle \sigma\left(  x\right)  \sigma\left(  y\right)  \right\rangle
_{c}=\left\langle x\left|  \frac{1}{-\partial^{2}+m_{\sigma}^{2}}\right|
y\right\rangle =\left\langle x\left|  K^{\left(  \sigma\right)  }\right|
y\right\rangle
\]%
\[%
\raisebox{-0cm}{\includegraphics[
height=0.3736in,
width=2.4993in
]%
{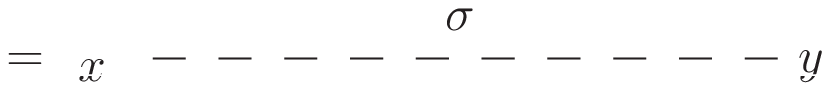}%
}%
\]%
\[
\left\langle \omega_{\mu}\left(  x\right)  \omega_{\nu}\left(  y\right)
\right\rangle _{c}=\left\langle x\left|  \frac{1}{\left(  -\partial
^{2}+m_{\omega}^{2}\right)  }\left(  \delta_{\mu\nu}-\frac{1}{m^{2}}%
\partial_{\mu}\partial_{\nu}\right)  \right|  y\right\rangle =\left\langle
x\left|  K_{\mu\nu}^{\left(  \omega\right)  }\right|  y\right\rangle
\]%
\[%
\raisebox{-0cm}{\includegraphics[
height=0.32in,
width=2.3108in
]%
{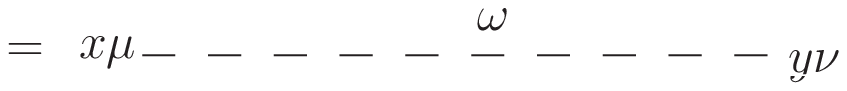}%
}%
\]
and:%
\[
\left\langle \pi_{a}\left(  x\right)  \pi_{b}\left(  y\right)  \right\rangle
_{c}=\delta_{ab}\left\langle x\left|  \frac{1}{-\partial^{2}+m_{\pi}^{2}%
}\right|  y\right\rangle =\delta_{ab}\left\langle x\left|  K^{\left(
\pi\right)  }\right|  y\right\rangle
\]%
\[%
\raisebox{-0cm}{\includegraphics[
height=0.3157in,
width=2.2295in
]%
{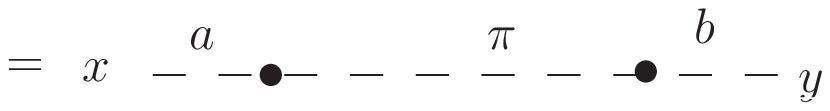}%
}%
\]%
\[
\left\langle \left(  \partial_{\mu}\pi_{a}\left(  x\right)  \right)  \pi
_{b}\left(  y\right)  \right\rangle _{c}=\delta_{ab}\left\langle x\left|
\partial_{\mu}\frac{1}{-\partial^{2}+m_{\pi}^{2}}\right|  y\right\rangle
=\delta_{ab}\left\langle x\left|  \partial_{\mu}K^{\left(  \pi\right)
}\right|  y\right\rangle
\]%
\[%
\raisebox{-0cm}{\includegraphics[
height=0.2966in,
width=2.1049in
]%
{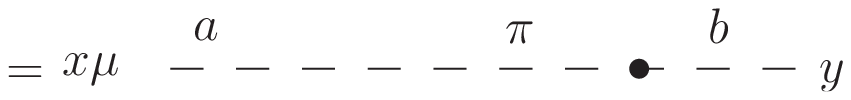}%
}%
\]%
\[
\left\langle \pi_{a}\left(  x\right)  \left(  \partial_{\nu}\pi_{b}\left(
y\right)  \right)  \right\rangle _{c}=-\delta_{ab}\left\langle x\left|
\frac{1}{-\partial^{2}+m_{\pi}^{2}}\partial_{\nu}\right|  y\right\rangle
=-\delta_{ab}\left\langle x\left|  K^{\left(  \pi\right)  }\partial_{\nu
}\right|  y\right\rangle
\]%
\[%
\raisebox{-0cm}{\includegraphics[
height=0.3918in,
width=2.4293in
]%
{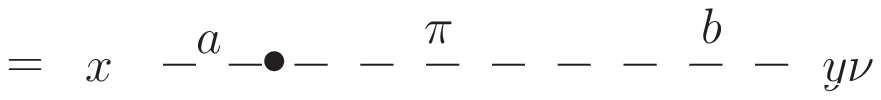}%
}%
\]%
\[
\left\langle \left(  \partial_{\mu}\pi_{a}\left(  x\right)  \right)  \left(
\partial_{\nu}\pi_{b}\left(  y\right)  \right)  \right\rangle _{c}%
=-\delta_{ab}\left\langle x\left|  \partial_{\mu}\frac{1}{-\partial^{2}%
+m_{\pi}^{2}}\partial_{\nu}\right|  y\right\rangle =-\delta_{ab}\left\langle
x\left|  \partial_{\mu}K^{\left(  \pi\right)  }\partial_{\nu}\right|
y\right\rangle
\]%
\begin{equation}%
{\includegraphics[
height=0.3797in,
width=2.6074in
]%
{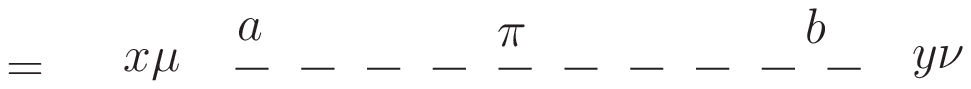}%
}%
\label{pimanb}%
\end{equation}
Expectation values of products of three or more fields can be expressed in
terms (\ref{omegaxmu}) and (\ref{pimanb}), using Wick's theorem.

Then $e^{-W_{0}\left(  j\right)  }e^{W\left(  j\right)  }$ is equal to the sum
of all distinct unlabeled diagrams formed by closed loops of oriented fermion
propagators joined to dashed boson propagator lines at points called
vertices.\ The contribution of a diagram is obtained by the following rules:

\begin{itemize}
\item  Assign a label $\sigma,\omega$ or $\pi$ to each dashed boson line .
Then assign distinct labels to the vertices.\ An example of a labeled diagram
is:%
\begin{equation}%
\raisebox{-0cm}{\includegraphics[
height=3.1419in,
width=3.9539in
]%
{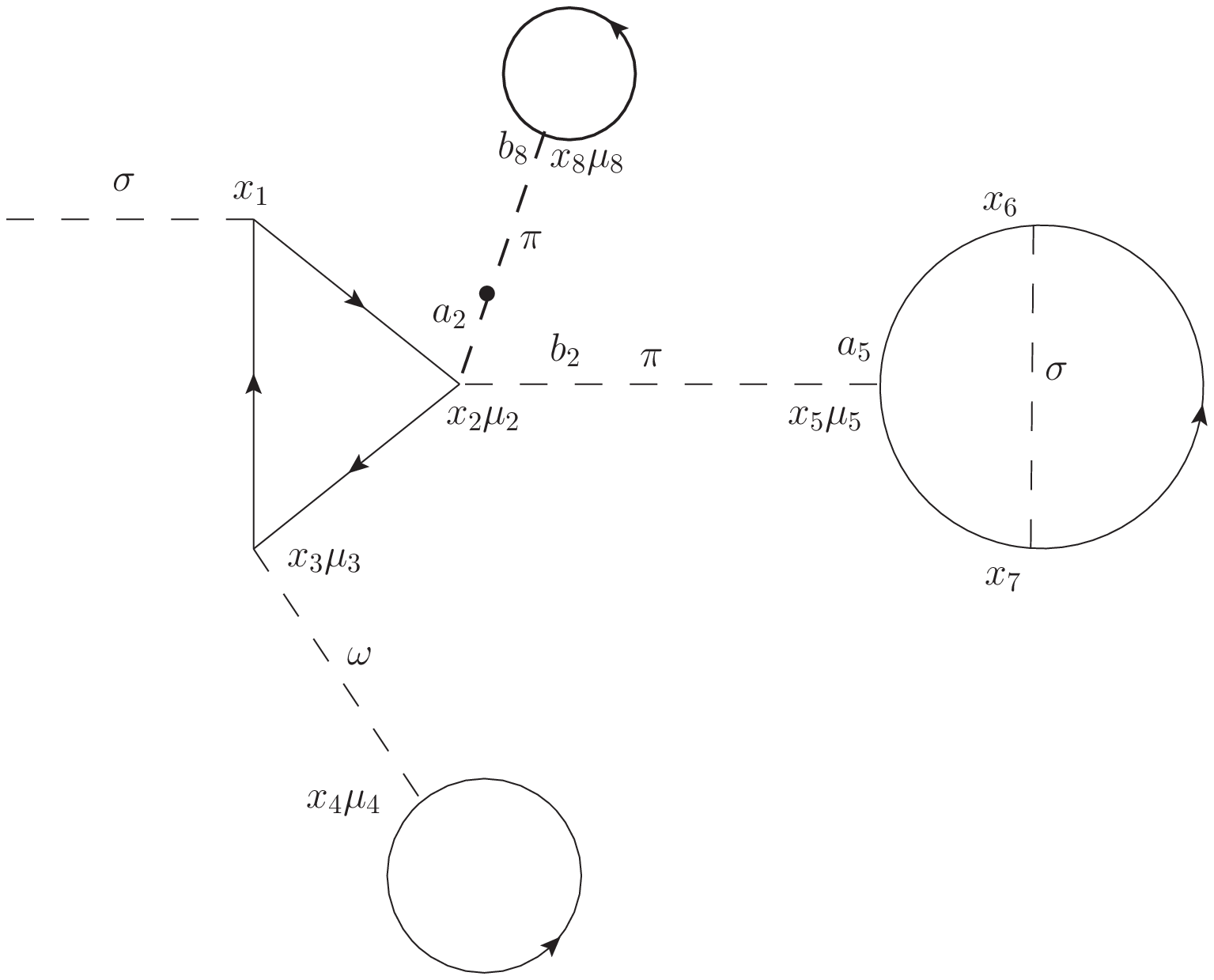}%
}%
\label{labl1}%
\end{equation}

\item  Each oriented fermion line, which reaches a point labeled $x_{1}$ and
which stems from a point labeled $x_{2}$, to which a meson dashed line (or two
meson lines) is attached, contributes one of the factors (\ref{twopivert}).

\item  Each meson dashed line contributes one of the factors (\ref{pimanb}) or
(\ref{omegaxmu}), depending on whether it joins one or two vertices. For
example, the two pion lines in the labelled diagram (\ref{labl1}), joining the
vertices bearing the labels $x_{2},x_{8}$ and $x_{5}$, contribute the
following factor (remember that the black dot indicates the \emph{absence} of
a derivative of the pion field):%
\begin{equation}
\left(  -\right)  \delta_{a_{2},b_{8}}\left\langle x_{2}\left|  \frac
{1}{-\partial^{2}+m_{\pi}^{2}}\partial_{\mu_{8}}\right|  x_{8}\right\rangle
\left(  -\right)  \delta_{b_{2},a_{5}}\left\langle x_{2}\left|  \partial
_{\mu_{2}}\frac{1}{-\partial^{2}+m_{\pi}^{2}}\partial_{\mu_{5}}\right|
x_{5}\right\rangle
\end{equation}

\item  The factors are written in the order in which they appear as one
follows each closed fermion loop and for each loop we take a trace $tr$ of
the product of fermion propagators forming the loop.

\item  A diagram containing $n_{v}$ vertices and $n_{l}$ closed loops is
multiplied by the factor $\left(  -\right)  ^{n_{v}+n_{l}}$. (Note that in the
absence of the source terms $j_{a}$, the number of vertices is always even.)

\item  Integrate over the space-time labels $x=\left(  \tau,\vec{r}\right)  $
of the vertices and divide by the symmetry factor, which is equal to the
number of permutations of the labels which lead to an identical unlabeled diagram.
\end{itemize}

The contributions of disconnected diagrams factor and their symmetry factors
are such that the sum of all diagrams is equal to the exponential of the sum
$\Gamma_{c}$ of connected diagrams. It follows that the partition function can
be expressed in terms of connected diagrams $\Gamma_{c}$ as follows:%
\begin{equation}
W\left(  j\right)  =W_{0}\left(  j\right)  +\Gamma_{c}%
\end{equation}
where $W_{0}\left(  j\right)  $ is given by (\ref{wzeroj}). In the absence of
the sources $j$, the partition function is given by:%
\begin{equation}
\ln\left(  Tre^{-\beta\left(  H-\mu N\right)  }\right)  =W=Tr\ln g^{-1}%
+\Gamma_{c}\label{conmes}%
\end{equation}
where $g$ is the fermion propagator (\ref{gprop}). An approximation to the
partition can be defined by a choice of a subset of connected diagrams.

\subsection{Diagram expansion of the particle densities.}

\label{ap:denmesnuc}

Let us add the following local source terms to $h_{0}$:%
\begin{equation}
\Gamma_{a}U_{a}\left(  x\right)  \equiv\Gamma^{\left(  \sigma\right)
}U_{\sigma}\left(  x\right)  +\Gamma_{\mu}^{\left(  \omega\right)  }U_{\mu
}\left(  x\right)  +\Gamma_{\mu a}^{\left(  \pi\right)  }U_{\mu a}\left(
x\right)  +\Phi_{a\left(  \mu b\right)  }^{\left(  \pi\right)  }U_{a\left(
\mu b\right)  }\left(  x\right)  \label{gauasrc}%
\end{equation}
From (\ref{partfun}) we see that upon a variation $U_{a}\left(  x\right)
\rightarrow U_{a}\left(  x\right)  +\delta U_{a}\left(  x\right)  $, we have:%
\begin{equation}
\delta\ln Z=\delta W=-\int d^{4}x\;\left\langle N^{\dagger}\left(  x\right)
\Gamma_{a}\delta U_{a}\left(  x\right)  N\left(  x\right)  \right\rangle
\end{equation}
so that:%
\begin{equation}
\rho_{a}\left(  x\right)  \equiv\left\langle N^{\dagger}\left(  x\right)
\Gamma_{a}N\left(  x\right)  \right\rangle =-\frac{\delta W}{\delta
U_{a}\left(  x\right)  } \label{rhoax1}%
\end{equation}
The operators $\Gamma_{a}$ are defined in (\ref{gamop}).\ They include the
coupling constants. Therefore the particle densities $n\left(  x\right)  $, in
the usual sense of the word, are related to the densities $\rho\left(
x\right)  $ as follows:%
\[
n_{\sigma}\left(  x\right)  \equiv\left\langle \bar{N}\left(  x\right)
N\left(  x\right)  \right\rangle =\frac{1}{g_{\sigma}}\rho_{\sigma}\left(
x\right)
\]%
\[
n\left(  x\right)  =\left\langle N^{\dagger}\left(  x\right)  N\left(
x\right)  \right\rangle =\frac{1}{ig_{\omega}}\rho_{\mu=0}\left(  x\right)
\]%
\[
n_{i}\left(  x\right)  =\left\langle N^{\dagger}\left(  x\right)  \alpha
_{i}N\left(  x\right)  \right\rangle =\frac{1}{g_{\omega}}\rho_{\mu=i}\left(
x\right)
\]%
\[
n_{\mu a}\left(  x\right)  =\left\langle N^{\dagger}\left(  x\right)
\beta\tau_{a}\gamma_{5}\gamma_{\mu}N\left(  x\right)  \right\rangle
=\frac{2f_{\pi}}{g_{A}}\rho_{\mu a}\left(  x\right)
\]%
\begin{equation}
n_{a\left(  \mu b\right)  }\left(  x\right)  =\left\langle \bar{N}\left(
x\right)  \varepsilon_{abc}\tau_{c}\pi_{a}\gamma_{\mu}N\left(  x\right)
\right\rangle =4f_{\pi}^{2}\rho_{a\left(  \mu b\right)  }\left(  x\right)
\label{nax}%
\end{equation}
(In the case of fermions with Coulomb interactions, only one density occurs
with $\Gamma_{a}=1$ so that $n\left(  x\right)  =\rho\left(  x\right)  $.)

In the presence of the source term, the fermion propagator (\ref{gprop})
becomes:%
\begin{equation}
g=\frac{1}{\partial_{\tau}+h_{0}-\mu+\Gamma_{a}U_{a}}\label{ggaua}%
\end{equation}
From the expression (\ref{conmes}) of the partition function, we see that the
particle densities (\ref{rhoax1}) can be expressed in terms of connected
diagrams as follows:%
\begin{equation}
\rho_{a}\left(  x\right)  \equiv\left\langle N^{\dagger}\left(  x\right)
\Gamma_{a}N\left(  x\right)  \right\rangle =-\left.  \frac{\delta}{\delta
U_{a}\left(  x\right)  }\left(  Tr\ln g^{-1}+\Gamma_{c}\right)  \right|
_{U=0}\label{rhodiag}%
\end{equation}
In (\ref{ggaua}), $\Gamma_{a}U_{a}$ is the following operator acting in the
Hilbert space of a single fermion:%
\begin{equation}
\Gamma_{a}U_{a}=\int d^{4}x\;\left|  x\right\rangle \Gamma_{a}U_{a}\left(
x\right)  \left\langle x\right|
\end{equation}
It follows that:%
\[
\frac{\delta}{\delta U_{a}\left(  x\right)  }\left\langle x_{1}\left|
g\right|  x_{2}\right\rangle =-\left\langle x_{1}\left|  g\right|
x\right\rangle \Gamma_{a}\left\langle x\left|  g\right|  x_{2}\right\rangle
\]%
\begin{equation}
\frac{\delta}{\delta U_{a}\left(  x\right)  }Tr\ln g^{-1}=Tr\;g\left|
x\right\rangle \Gamma_{a}\left\langle x\right|  =tr\;\left\langle x\left|
g\right|  x\right\rangle \Gamma_{a}\label{dgduxa}%
\end{equation}
The first term of (\ref{rhodiag}) is the unperturbed particle density:%
\begin{equation}
\rho_{a}^{\left(  0\right)  }\left(  x\right)  =-tr\;\left\langle x\left|
g\right|  x\right\rangle \Gamma_{a}%
\end{equation}
We can use (\ref{g12zerot}) to get:%
\begin{equation}
\rho_{a}^{\left(  0\right)  }\left(  x\right)  =\sum_{h}\left\langle \left.
h\right|  \vec{r}\right\rangle \Gamma_{a}\left\langle \left.  \vec{r}\right|
h\right\rangle =\rho^{\left(  0\right)  }\left(  \vec{r}\right)
\end{equation}
For the second term of (\ref{rhodiag}) we note that the potential $U_{a}$
occurs only in the propagators $g$ of the connected diagrams $\Gamma_{c}$ so
that:%
\begin{equation}
\frac{\delta\Gamma_{c}}{\delta U_{a}\left(  x\right)  }=\int d^{4}x_{1}%
d^{4}x_{2}\;\frac{\delta\Gamma_{c}}{\delta\left\langle x_{1}\left|  g\right|
x_{2}\right\rangle }\frac{\delta\left\langle x_{1}\left|  g\right|
x_{2}\right\rangle }{\delta U_{a}\left(  x\right)  }=-\int d^{4}x_{1}%
d^{4}x_{2}\;\frac{\delta\Gamma_{c}}{\delta\left\langle x_{1}\left|  g\right|
x_{2}\right\rangle }\left\langle x_{1}\left|  g\right|  x\right\rangle
\Gamma_{a}\left\langle x\left|  g\right|  x_{2}\right\rangle
\end{equation}
The particle density can therefore be expressed in terms of the connected
diagrams thus:%
\begin{equation}
\rho_{a}\left(  x\right)  =-tr\;\left\langle x\left|  g\right|  x\right\rangle
\Gamma_{a}+\int d^{4}x_{1}d^{4}x_{2}\;\frac{\delta\Gamma_{c}}{\delta
\left\langle x_{1}\left|  g\right|  x_{2}\right\rangle }\left\langle
x_{1}\left|  g\right|  x\right\rangle \Gamma_{a}\left\langle x\left|
g\right|  x_{2}\right\rangle \label{rhoax}%
\end{equation}
The second term is the sum of distinct diagrams obtained from the unlabeled
connected diagrams $\Gamma_{c}$ by inserting a slash onto its oriented fermion
lines, exactly as in (\ref{retdiag}), except that the slash now bears the
label $\left(  x,a\right)  $. \ An oriented fermion line, stemming from a
vertex labelled $x_{2}$, reaching a vertex labelled $x_{1}$, and bearing a
slash labelled $\left(  x,a\right)  $ contributes a factor equal to:%
\begin{equation}%
{\includegraphics[
height=1.1744in,
width=2.4924in
]%
{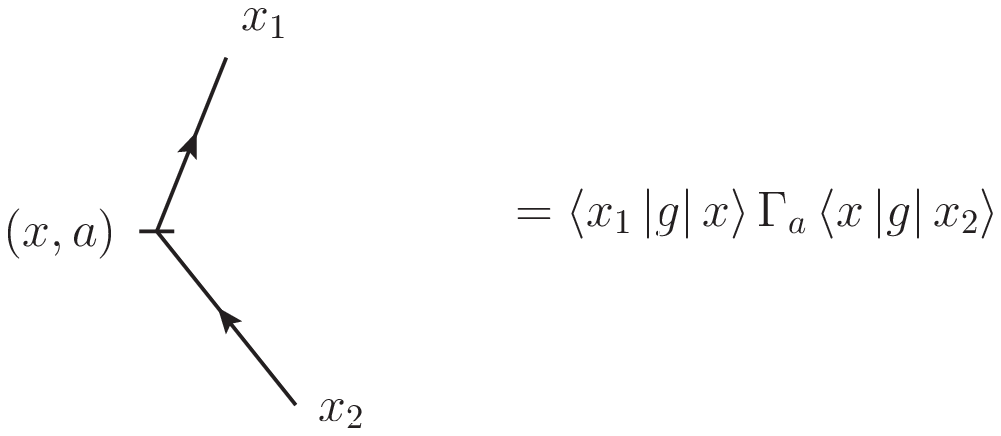}%
}%
\label{slashxa}%
\end{equation}
The slash does not modify the sign of the diagram.

\renewcommand{\theequation}{\Alph{section}.\arabic{equation}}

\setcounter{equation}{0}

\section{Particle and hole orbits and density-density correlation functions.}

When $h_{0}$ is time-independent, the unperturbed fermion propagator
$g^{-1}=\left(  \partial_{\tau}+h_{0}-\mu\right)  $ is diagonal in the
$\left|  \omega\lambda\right\rangle $ representation, where the states
$\left|  \lambda\right\rangle $ are eigenstates of $h_{0}$:
\begin{equation}
h_{0}\left|  \lambda\right\rangle =e_{\lambda}\left|  \lambda\right\rangle
\label{hzerolam}%
\end{equation}
and where $\left|  \omega\right\rangle $ are the (euclidean) time plane wave
states:
\begin{equation}
\left\langle \left.  \tau\right|  \omega\right\rangle =\frac{1}{\sqrt{\beta}%
}e^{i\omega\tau}\;\;\;\;\;\;\;\partial_{\tau}\left|  \omega\right\rangle
=i\omega\left|  \omega\right\rangle
\end{equation}
so that:
\begin{equation}
g^{-1}\left|  \omega\lambda\right\rangle =\left(  \partial_{\tau}+h_{0}%
-\mu\right)  \left|  \omega\lambda\right\rangle =\left(  i\omega+e_{\lambda
}-\mu\right)  \left|  \omega\lambda\right\rangle \label{lamzero}%
\end{equation}

In the zero temperature limit $\beta\rightarrow\infty$, the matrix elements of
$g$ are equal to:
\[
\left\langle x_{1}\left|  g\right|  x_{2}\right\rangle \equiv\left\langle
\tau_{1}\vec{r}_{1}\left|  g\right|  \tau_{2}\vec{r}_{2}\right\rangle
=\sum_{\omega\lambda}\left\langle \left.  \tau_{1}\vec{r}_{1}\right|
\omega\lambda\right\rangle \frac{1}{i\omega+e_{\lambda}-\mu}\left\langle
\left.  \omega\lambda\right|  \tau_{2}\vec{r}_{2}\right\rangle
\]%
\begin{equation}
=\theta\left(  \tau_{1}-\tau_{2}\right)  \sum_{p}e^{-\left(  e_{p}-\mu\right)
\left(  \tau_{1}-\tau_{2}\right)  }\left\langle \left.  \vec{r}_{1}\right|
p\right\rangle \left\langle \left.  p\right|  \vec{r}_{2}\right\rangle
-\theta\left(  \tau_{2}-\tau_{1}\right)  \sum_{h}e^{-\left(  e_{h}-\mu\right)
\left(  \tau_{1}-\tau_{2}\right)  }\left\langle \left.  \vec{r}_{1}\right|
h\right\rangle \left\langle \left.  h\right|  \vec{r}_{2}\right\rangle
\label{g12}%
\end{equation}
where the ''particle'' and ''hole'' orbits $\left|  p\right\rangle $ and
$\left|  h\right\rangle $ are the eigenstates $\left|  \lambda\right\rangle $
belonging to energies respectively $e_{p}>\mu$ and $e_{h}<\mu$. The equal time
propagator is:%
\begin{equation}
\left\langle \tau\vec{r}_{1}\left|  g\right|  \tau\vec{r}_{2}\right\rangle
=-\sum_{h}\left\langle \left.  \vec{r}_{1}\right|  h\right\rangle \left\langle
\left.  h\right|  \vec{r}_{2}\right\rangle \label{g12zerot}%
\end{equation}
We shall encounter the so-called density-density correlation function:%
\[
\left\langle \vec{r}_{1}\left|  D\right|  \vec{r}_{2}\right\rangle
\equiv-tr\int d\tau\left\langle \tau_{1}\vec{r}_{1}\left|  g\right|  \tau
\vec{r}\right\rangle \left\langle \tau\vec{r}\left|  g\right|  \tau_{1}\vec
{r}_{2}\right\rangle =-tr\int d\tau_{1}\left\langle \tau_{1}\vec{r}_{1}\left|
g\right|  \tau\vec{r}\right\rangle \left\langle \tau\vec{r}\left|  g\right|
\tau_{1}\vec{r}_{2}\right\rangle
\]%
\[
=\sum_{ph}\frac{1}{e_{p}-e_{h}}\left(  \left\langle \left.  p\right|  \vec
{r}_{2}\right\rangle \left\langle \left.  \vec{r}_{2}\right|  h\right\rangle
\left\langle \left.  h\right|  \vec{r}_{1}\right\rangle \left\langle \left.
\vec{r}_{1}\right|  p\right\rangle +\left\langle \left.  h\right|  \vec{r}%
_{2}\right\rangle \left\langle \left.  \vec{r}_{2}\right|  p\right\rangle
\left\langle \left.  p\right|  \vec{r}_{1}\right\rangle \left\langle \left.
\vec{r}_{1}\right|  h\right\rangle \right)
\]%
\begin{equation}%
\raisebox{-0cm}{\includegraphics[
height=1.2047in,
width=1.4295in
]%
{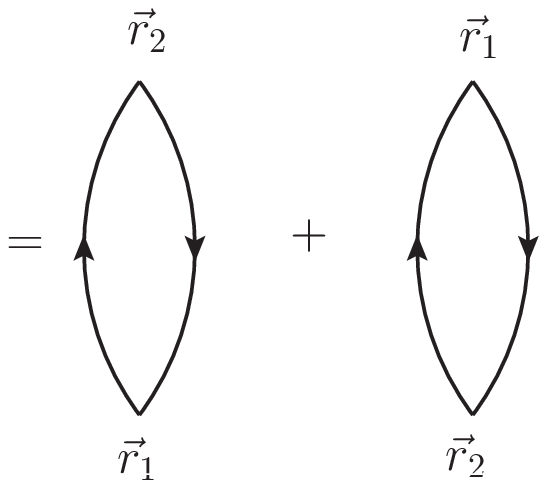}%
}%
\label{drr}%
\end{equation}
The diagram (\ref{drr}) is a time-independent Goldstone diagram in which
upgoing propagators are particle orbits and downgoing propagators hole orbits.
The \emph{inverse} correlation function will be represented in diagrams by a
double line:%
\[
\left\langle \vec{r}_{1}\left|  D^{-1}\right|  \vec{r}_{2}\right\rangle
\]%
\begin{equation}%
\raisebox{-0cm}{\includegraphics[
height=0.269in,
width=1.7962in
]%
{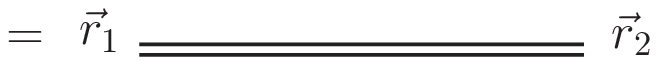}%
}%
\label{drrm1}%
\end{equation}

\section*{Acknowledgments}

I thank Jean Paul Blaizot for reading the manuscript and for helpful 
suggestions. I also thank Thomas Duguet and Bertrand Giraud for 
instructive discussions. The Feynman diagrams were drawn with 
the program Jaxodraw.

\bibliographystyle{unsrt}
\bibliography{Njl}
\end{document}